\documentclass[12pt,preprint]{aastex}

\def\AaA{{\em Astr.~Astrophys.}}

\def\ApJ{{\em Astrophys.~J.}}

\def\ApJS{{\em Astrophys.~J.~Suppl.}}

\def\ASPC{{\em ASP~Conf.~Series}}

\def\CPC{{\em Comp.~Phys.~Communic.}}
\def\MN{{\em Mon.~Not.~R.~astr.~Soc.}}
\def\Nat{{\em Nature}}

\def\PR{{\em Phys.~Reports}}

\def\etal{{et al.\thinspace}}

\def\spose#1{\hbox to 0pt{#1\hss}}

\def\multleft#1{\hbox to size{\vbox {\halign {\lft{##}\cr #1}}\hfill}\par}
\def\multright#1{\hbox to size{\vbox {\halign {\rt{##}\cr #1}}\hfill}\par}

\def\degmark{^\circ}
\def\boxit#1{\vbox{\hrule\hbox{\vrule\kern3pt\vbox{\kern3pt
          #1 \kern3pt}\kern3pt\vrule}\hrule}}

\def\cm{{\rm\thinspace cm}}

\def\erg{{\rm\thinspace erg}}
\def\eV{{\rm\thinspace eV}}

\def\GHz{{\rm\thinspace GHz}}

\def\keV{{\rm\thinspace keV}}
\def\km{{\rm\thinspace km}}

\def\Mpc{{\rm\thinspace Mpc}}
\def\Msun{\hbox{$\rm\thinspace M_{\odot}$}}

\def\ph{{\rm\thinspace ph}}
\def\s{{\rm\thinspace s}}
\def\ks{{\rm\thinspace ks}}

\def\ergpcmps{\hbox{$\erg\cm^{-1}\s^{-1}\,$}}

\def\ergps{\hbox{$\erg\s^{-1}\,$}}

\def\kmps{\hbox{$\km\s^{-1}\,$}}

\def\pcmsq{\hbox{$\cm^{-2}\,$}}

\def\phpcmsqps{\hbox{$\ph\cm^{-2}\s^{-1}\,$}}

\makeatletter

\let\@internalcite\cite
\def\cite{\@ifstar{\citey}{\citefull}}
\def\citefull{\def\astroncite##1##2{##1\ ##2}\@internalcite}
\def\citey{\def\astroncite##1##2{##1\ (##2)}\@internalcite}
\def\citeyear{\def\astroncite##1##2{##2}\@internalcite}
\def\citename{\def\astroncite##1##2{##1}\@internalcite}
\def\@citex[#1]#2{\if@filesw\immediate\write\@auxout{\string\citation{#2}}\fi
  \def\@citea{}\@cite{\@for\@citeb:=#2\do
    {\@citea\def\@citea{; }\@ifundefined
       {b@\@citeb}{{\bf ??}\@warning
       {Citation `\@citeb' on page \thepage \space undefined}}%
{\csname b@\@citeb\endcsname}}}{#1}}
 
\def\@cite#1#2{#1\if@tempswa #2\fi} 
\def\@biblabel#1{}
 
\def\astroncite#1#2{#1\ #2}
\makeatother

\begin{document}

\title{Constraining Black Hole Spin Via X-ray Spectroscopy}

\author{Laura~W.~Brenneman\altaffilmark{1}, 
Christopher~S.~Reynolds\altaffilmark{1}}

\altaffiltext{1}{Dept. of Astronomy, University of Maryland, College
Park, College Park MD~20742}

\begin{abstract}
We present an analysis of the observed broad iron line feature and
putative warm absorber in the long 2001 {\it XMM-Newton} observation
of the Seyfert-1.2 galaxy MCG--6-30-15.  The new {\tt kerrdisk} model
we have designed for simulating line emission from accretion disk
systems allows black hole spin to be a free parameter in the fit,
enabling the user to formally constrain the angular momentum of a
black hole, among other physical parameters of the system.  In an
important extension of previous work, we derive constraints on the
black hole spin in MCG--6-30-15 using a self-consistent model for
X-ray reflection from the surface of the accretion disk while
simultaneously accounting for absorption by dusty photoionized
material along the line of sight (the warm absorber).  Even including
these complications, the {\it XMM-Newton}/EPIC-pn data require extreme
relativistic broadening of the X-ray reflection spectrum; assuming no
emission from within the radius of marginal stability, we derive a
formal constraint on the dimensionless black hole spin parameter of
$a=0.989^{+0.009}_{-0.002}$ at $90\%$ confidence.  The principal
unmodeled effect that can significantly reduce the inferred black hole
spin is powerful emission from within the radius of marginal
stability.  Although significant theoretical developments are required
to fully understand this region, we argue that the need for a rapidly
spinning black hole is robust to physically plausible levels of
emission from within the radius of marginal stability.  In particular,
we show that a non-rotating black hole is strongly ruled out.
\end{abstract}


\section{Introduction}
\label{sec:intro}

Accreting black holes are the driving force behind some of the most
powerful processes in the universe.  The central regions of active
galactic nuclei (AGN), in particular, are prodigious sources across
the electromagnetic spectrum including the X-ray band. The X-ray
continuum (which is accurately approximated by a power-law up to $\sim
100 \keV$) is thought to be produced by Comptonization processes in
the corona (or base of a jet) surrounding the inner part of the
accretion disk.  A portion of the X-ray photons produced in the corona
irradiate the underlying optically-thick disk, producing the
so-called ``X-ray reflection'' signatures in the observed spectrum
(Guilbert \& Rees 1988; Lightman \& White 1988).  These X-ray
reflection signatures consist of fluorescent and recombination
emission lines sitting on a continuum due to Compton scattering and
the summed radiative recombination continua of the excited ions in the
photoionized disk surface (George \& Fabian 1991; Ross \& Fabian
2005).  The Fe-K$\alpha$ line is the most prominent of these features
due to its energy (at $6.4 \keV$ it is visible above the direct power
law continuum), and the high astrophysical abundance and fluorescent
yield of iron.  Additionally, this line is often significantly
broadened by both the standard Doppler effect and relativistic 
processes, the effects of
which increase the closer the line is emitted to the event horizon
(Fabian \etal 1989; Laor 1991).  These relativistic processes include
light bending, beaming and gravitational redshifting, all of which
result in a greatly elongated and skewed line profile; in particular,
the line profile can display an extended low-energy tail primarily
resulting from gravitational redshift.  The line emission region is
sufficiently close to the black hole that frame-dragging effects
associated with the black hole spin can be important in determining
the line profiles.  The broad iron line is therefore a powerful probe
of the relativistic effects on the spacetime immediately surrounding
the black hole.

The first broad iron line robustly detected and resolved in an AGN was
found by the {\it Advanced Satellite for Cosmology and Astrophysics
(ASCA)} in the Seyfert-1.2 galaxy MCG--6-30-15 (Tanaka \etal 1995;
Iwasawa \etal 1996), and since then has been extensively studied with
{\it BeppoSAX} (Guainazzi \etal 1999), {\it RXTE} (Lee \etal 1999,
2000), {\it Chandra} (Lee \etal 2002; Young \etal 2005) and {\it
XMM-Newton} (Wilms \etal 2001; Fabian \etal 2002; Reynolds \etal
2004).  All of these results show that the broad iron line feature is
consistent with a highly redshifted line from the inner parts of an
accretion disk; no alternative hypothesis has yet explained the
spectrum of MCG--6-30-15 satisfactorily (Fabian \etal 1995; Reynolds
\& Wilms 2000; Vaughan \& Fabian 2004; Young \etal 2005).  Subsequent
{\it ASCA}, {\it Chandra} and {\it XMM-Newton} studies have discovered
similarly broad iron line profiles in several other Seyferts, such as
MCG--5-23-16 (Dewangen, Griffiths \& Schurch 2003), NGC 3516 (Turner
\etal 2002), Mrk335 (Gondoin \etal 2002), and Mrk766 (Pounds \etal
2003a).

In addition to having a sample of objects that have been observed with
robust broad iron lines, it is equally important to have a precise
model to use in fitting the data.  The two line profiles currently
included as standard in the current versions of the commonly used
spectral fitting package {\sc xspec} (Arnaud 1996) are useful as a
starting point, but ultimately quite limited in terms of their ability
to accurately parameterize the line.  The {\tt diskline} model (Fabian
\etal 1989) describes the line profile from a disk around a
non-rotating (Schwarzschild) black hole, and, due to the
approximations employed, does not include relativistic light bending.
Similarly, the {\tt laor} model (Laor 1991) has important constraints
as well: this is a fully-relativistic model, but the dimensionless
spin parameter of the black hole is hard-wired at $a=0.998$, the
equilibrium spin of a black hole accreting from a Novikov \& Thorne
(1974) accretion disk. Furthermore, due to the computational realities
of the early 1990s, the relativistic transfer functions underlying the
{\tt laor} model are pre-calculated and tabulated, yielding noise (or
even gross inaccuracies) in the line profiles produced, especially at
very high disk inclination angles.  Given these limitations, as well
as the high quality of AGN spectra currently being obtained with {\it
Chandra} and {\it XMM-Newton} (and hopefully in the future with the
remaining operable instruments aboard {\it Suzaku}), it is imperative
that X-ray astronomers have access to effective models that are fully
relativistic, accurate, and that allow black hole spin to be fit as a
free parameter.  This latter point is crucial if we are to hope to
constrain the spin of astrophysical black holes using broad iron lines
and other X-ray reflection signatures.

Three new relativistic line models have recently been developed for
this purpose and implemented in a form that can be readily used by
X-ray astronomers: the {\tt ky} suite (Dov\v{c}iak \etal 2004) and similar
codes by Beckwith \& Done (2004) and \v{C}ade\v{z} \& Calvani (2005).  These
models achieve comparable results for the morphologies of the line
profiles, and all offer significant improvements over the
{\tt diskline} and {\tt laor} results in terms of accuracy and
precision over a wider range of physical parameters.  Most
importantly, these models leave the spin of the black hole as a free
parameter, include emission from within the innermost stable circular
orbit in the disk, and compute fully relativistic photon transfer
functions.  In practice, the {\sc xspec} modules that implement both
the {\tt ky} and Beckwith \& Done models use relativistic transfer
functions stored in very large (multi-gigabyte) pre-calculated tables.
Using these models, both groups conclude that fitting broad iron lines
cannot truly constrain the spin of the black hole.  Since their models
allow emission from {\it any} radius of the accretion disk outside of
the event horizon, they can produce iron line profiles with
arbitrarily redshifted wings even if the underlying black hole
spacetime has no spin.  One must consider the physicality of this
assumption, however: it is not possible to get line emission from
radii deep within the plunging region for several reasons.  Firstly,
the closer one gets to the event horizon, the smaller the geometrical
area of the disk available for emission becomes, while the
gravitational redshift affecting any emission from such regions
increases correspondingly.  Both of these effects would tend to
minimize the contribution to the line profile from most of the emission 
inside the radius of marginal stability.
Further, the ionization of the disk material within $\sim 4 \,r_{\rm g}$
of a Schwarzchild black hole is inevitably too high to produce
significant line emission in the first place, and the optical depth of
this material also decreases rapidly within the radius of marginal
stability (Reynolds \& Begelman 1997; Young \etal 1998).  
For all of these reasons, one
simply cannot produce significant line emission from any arbitrary
radius outside the event horizon.  This must be taken into account
when modeling such line emission.

In this paper, we undertake a rigorous re-analysis of the broad iron
line seen in the $\sim 350 \ks$ July-2001 {\it XMM-Newton} observation
(Fabian \etal 2002).  We assess the constraints on the iron line
profile and black hole spin including the complications introduced by
other spectral components displayed by this system, especially the
substantial column of absorbing photoionized gas seen along the line
of sight to the central disk of this AGN.  To facilitate this
investigation, we present a new variable-spin accretion disk line
profile model, {\tt kerrdisk}, that we are currently developing for
public use in the {\sc xspec} package.  The principal difference
between {\tt kerrdisk} and the previously mentioned variable-spin line
profile models is that our model does not rely on gigabyte-sized,
precomputed tables of the photon transfer function over many thousands
of radii within the disk.  Rather, as we will discuss below, we use
the approach of Cunningham (1975) to decompose the line profile
integral into a simple analytic component (describing the effects of
relativistic beaming) and an additional transfer function (describing
the effects of light bending) which is a slowly varying function of
black hole spin and disk inclination angle and hence only needs to be
coarsely tabulated. This technique allows us to obtain high-quality
line profiles with very modest sized (40\,MB) tables, making it
substantially more portable and convenient than some other models.
This accuracy, portability and comparative speed of computation make
{\tt kerrdisk} a valuable addition to the X-ray astronomer's software
arsenal.  Additionally, a trivial change of the driver code allows the
user to implement any desired limb-darkening law or compute the line
profiles to arbitrary resolution and accuracy (if, for example, the
user is interested in the narrow cusps of the line profile).  Neither
of these modifications is possible if the limb-darkening and
resolution are hard-wired into the tabulated relativistic transfer
functions.

In \S2, we motivate our study by giving a brief recap of the
history of the broad iron line in MCG--6-30-15.  \S3 describes
our new iron line model, including a detailed comparison of {\tt
kerrdisk} to the other line profile models already discussed.
\S4 presents our analysis of MCG--6-30-15 using these
variable-spin line models.  This work is performed under the
``standard'' assumption that there is no observable X-ray reflection
from inside the radius of marginal stability of the accretion disk.  A
future paper will extend the analysis to the more general case
where physically reasonable levels of emission from this region are
allowed.  Under this assumption we arrive at the conclusion that the black
hole in MCG--6-30-15 is a rapid-rotator.  \S5 and \S6 discuss our
results and draw conclusions, respectively.

\section{A Brief History of Broad Iron Line Studies in MCG--6-30-15}
\label{sec:history}

MCG--6-30-15 is an S0-type galaxy in the constellation of Centaurus
that hosts a Seyfert-1.2 nucleus.  It has a measured redshift of $z =
0.008$, placing it at a distance of $d \approx 37 \Mpc$ using WMAP
cosmological parameters.  X-ray studies of this AGN have revealed a
powerful central source along with a significantly broadened
Fe-K$\alpha$ feature.  This broad line has led MCG--6-30-15 to become
one of the most studied AGN in the X-ray band due to its potential
as a probe of black hole and accretion disk physics.  Although the
mass of the hole is not yet well constrained, estimates based on X-ray
studies have placed it
in the range of $10^6-2 \times 10^7 \Msun$ (Nowak \& Chiang 2000;
Reynolds 2000).  
McHardy \etal (2005) have further narrowed this mass range
to $3-6 \times 10^6 \Msun$ using a variety of methods such as the
M-$\sigma$ relation, line widths from optical spectra, BLR
photoionization arguments and X-ray variability.

Reflection signatures from neutral material had been
observed with {\it EXOSAT} (Nandra \etal 1989) and {\it Ginga}
(Nandra, Pounds \& Stewart 1990; Matsuoka \etal 1990), but it was not
until deep observations of MCG--6-30-15 by {\it ASCA} that the
broadened and skewed iron line was robustly detected (Tanaka \etal
1995).  It was determined that the line profile matched that expected
due to X-ray reflection from the surface of a relativistic accretion
disk; the robustness of this interpretation was demonstrated in Fabian
\etal (1995).  Broad-band {\it BeppoSAX} data taken by Guainazzi \etal
(1999) confirmed the Tanaka \etal detection of a broadened and
redshifted iron line with an equivalent width of ${\rm EW} \approx 200
\eV$.

A detailed re-analysis of this observation by Iwasawa \etal (1996)
identified a period of time when the source entered the so-called
``deep minimum'' state, marked by low continuum emission.  While in this
state, the iron line width (and especially the extent of
the red-wing of the line) markedly increased to the point that
emission from a disk around a Schwarzschild black hole truncated at
the radius of marginal stability ($r_{\rm ms}=6 \,r_{\rm g}$) could no
longer reproduce the observed line profile.  Noting that the radius of
marginal stability moves inwards (to a location characterized by
higher gravitational redshift) as the black hole spin is increased, it
was subsequently argued that the black hole in MCG--6-30-15 must be
rapidly rotating.  Fitting sequences of Novikov \& Thorne (1974)
models to the deep minimum {\it ASCA} data (and assuming that the
X-ray irradiation tracks the local dissipation in the underlying
disk), Dabrowski \etal (1997) derived a lower limit of $a>0.94$ on
the rotation of the black hole.  At this point, however, Reynolds \&
Begelman (1997) noted that physically plausible scenarios could result
in sufficient X-ray reprocessing (including ionized iron line
emission) from within the radius of marginal stability to fit the {\it
ASCA} deep minimum state with even a Schwarzschild black hole.  In
order to illuminate the region of the disk within $r_{\rm ms}$, they
hypothesized an X-ray source on the symmetry axis some height above
the disk, and suggested that iron line profile changes (and some part
of the X-ray continuum flux changes) could be attributed simply to
changes in the height of this X-ray source.

After the discovery of its broad iron line feature, MCG--6-30-15
became the subject of many more observations.  With the first
observations of {\it XMM-Newton} in 2000, astronomers had a new tool
with unparalleled throughput with which to examine this source in
finer detail.  The first {\it XMM-Newton} observation of MCG--6-30-15
was fortunate enough to catch the source in its ``deep minimum'' state
characterized by low continuum flux and a broader than normal
Fe-K$\alpha$ profile.  The resulting high signal-to-noise spectrum
revealed an extremely extended red-wing to the line profile extending
down to $3-4 \keV$ (Wilms \etal 2001).  Because so much emission
seemed to be coming from radii deep within the gravitational potential
well, and because the emissivity index was correspondingly quite high,
the authors hypothesized an interaction between the spinning black hole
and its accretion disk.  Magnetic torquing effects between the two
could result in the extraction of rotational energy from the black
hole which would, in turn, power the high coronal emission seen from
the inner radii of the disk (Agol \& Krolik 2000; Garofalo \& Reynolds
2005).  An alternative mechanism consists of the strong gravitational
focusing of a high-latitude source above a rapidly rotating black hole
(Martocchia \& Matt 1996; Miniutti \& Fabian 2004).  By contrast with
the {\it ASCA} era, a Schwarzschild model appeared not to work for
these data --- a line extending down to $\sim 3 \keV$ would require
line emission from extremely deep within the plunge region: $r\approx
3 \,r_{\rm g}=0.5 \,r_{\rm ms}$, a situation which appears unphysical due
to the high ionization expected in this part of the flow (Reynolds \&
Begelman 1997).

Reynolds \etal (2004) performed a follow-up and more detailed analysis of
the {\it XMM-Newton} observation of MCG--6-30-15 taken by Wilms \etal
(2001).  They explicitly demonstrated that the iron line profile was
inconsistent with an X-ray irradiation profile that follows a Novikov
\& Thorne (1974) dissipation law, even for an extremal Kerr black
hole.  Employing the generalized thin disk model of Agol \& Krolik
(2000) that includes a torque applied at $r=r_{\rm ms}$, Reynolds et
al. (2004) suggested that this torque has a great impact on the disk
emission seen in this observation.  In the ``deep minimum'' state, the
torqued-disk scenario requires that the disk is largely emitting via
the extraction of rotational energy from the black hole rather than
via accretion.  These authors also examine spectral variability during
the deep minimum state.  By examining difference spectra and direct
spectral fits to $10 \ks$ segments of data from this observation, the
authors concluded that the intensity of the broad line seems to be
proportional to the hard $2-10 \keV$ flux of the source: the
equivalent width of the line remains approximately constant while the
source fluctuates substantially in amplitude.  Such behavior is
consistent with simple X-ray reflection models.

The longest {\it XMM-Newton} observations of the source to date was
published by Fabian \etal (2002).  This group found MCG--6-30-15 in
its normal state, and recorded data for over $87$ hours.  In this
state, the bulk of the iron line emission was in a narrower line
compared with the ``deep minimum'' state, although a very extended
red-wing was still evident.  The time-averaged EPIC-pn spectrum again
showed that the Fe-K$\alpha$ feature was very strong, and the long
data set was of sufficient resolution and quality that the spectrum
demanded a fit incorporating a full reflection model.  Relativistic
smearing needed to be applied not just to the cold iron line, but to
the entire reflection continuum.  Taking this into account, the iron
line was once again found to produce emission within the radius of
marginal stability for a Schwarzchild black hole.  Fitting the line
with the maximally spinning ($a=0.998$) black hole {\tt laor} model
suggested that the line emissivity followed a $\epsilon \propto
r^{-\alpha}$ dependence with $\alpha=4.5-6$ between an inner radius
$r_{\rm min}<2\,r_{\rm g}$ and a break radius $r_{\rm br} \approx
6\,r_{\rm g}$.  Beyond the break radius, the emissivity profile
flattened to $\alpha \sim 2.5$.  This broken power-law form was
strongly preferred by the data over the usual simple power-law
emissivity functions usually fitted to such data --- again, this
reflects the high quality of the data.  Fabian \etal (2002) also note
that, in its normal state, difference spectra of MCG--6-30-15 show
spectral variability from $2-10 \keV$ in the form of a power-law: the
iron line flux changed little between successive $10 \ks$ frames of
the observation, whereas the continuum flux varied by as much as a
factor of $\sim 2$.  This result is in contrast to that found by
Reynolds \etal (2004), who observed an iron line flux proportional to
the $2-10 \keV$ continuum flux when MCG--6-30-15 was in its ``deep
minimum'' state.  Comparing these two studies, it appears that the
iron line flux is proportional to the observed X-ray power-law
continuum at low fluxes and then ``saturates'' to an approximately
constant level once the observed X-ray continuum exceeds a certain
level.  This complexity could be due to light-bending effects if the
power-law X-ray continuum source is situated close to the spin axis of
the black hole (Miniutti \& Fabian 2004).  Alternatively, patchy
ionization of the disk surface might produce such a saturation
(Reynolds 2000).

Dov\v{c}iak \etal (2004) utilize their {\tt ky} suite of iron line
profile models to fit the time-averaged spectrum of MCG--6-30-15 as
well, also using data from the long {\it XMM-Newton} observation of
Fabian \etal (2002).  They fit the $3-10 \keV$ spectrum with four
combinations of {\tt ky} models, all involving a broad Fe-K$\alpha$
{\tt kyrline}, a narrow Gaussian emission line at $6.9 \keV$ (likely
an ionized line of iron, as cited in Fabian \etal (2002)), and a
Compton-reflection continuum from a relativistic disk (smeared using a
{\tt kyconv} kernel).  The authors also find that models describing
the disk emissivity as a broken power-law rather than a single
power-law in radius achieve significantly better statistical fits,
though among these broken power-law models, comparable fits can be
obtained for a wide variety of black hole spin values.  These authors
therefore conclude that iron line profiles are not a good way to
constrain black hole spin.  At some level, this objection amounts to
obvious point that arbitrarily large redshifts can be obtained around
{\it any} black hole (rotating or not) if one is at liberty to produce
emission from any radius arbitrarily close to the event horizon.  One
must, however, consider the physical realism of this assumption in
general, and the best-fit spectral parameters determined by these
authors in particular.  For their low-spin ($a=0.25$) best-fitting
model, Dov\v{c}iak \etal concluded that a substantial amount of iron
line emission had to originate from deep within the radius of marginal
stability; translating the fit parameters of Dov\v{c}iak \etal into
standard quantities, they require an inner emissivity profile of
$\epsilon \propto r^{-9}$ starting at an inner radius of $r_{\rm
in}=3.2\,GM/c^2$, to be compared with the radius of marginal stability
for a $a=0.25$ black hole which is at $r_{\rm ms} \approx
5.2\,GM/c^2$.  As has already been noted and will be explored further
in \S4, it is hard to understand how this region of the accretion flow
could contribute to any part of the observed iron emission given that
the fact it will be very tenuous and extremely highly photoionized
(Reynolds \& Begelman 1997; Young, Ross \& Fabian 1998).

The medium resolution (CCD) data discussed so far leave ambiguous the
possible role of complex ionized absorption in distorting the observed
X-ray continuum shape and hence the inferred iron line profile.  To
assess the role of this absorption, Young \etal (2005) performed and
analyzed a deep, $522 \ks$ grating observation of MCG--6-30-15 in
May-2004 with the HETGS instrument on {\it Chandra}.  This observation
produced two important results: first, the authors found that the
difference in the hard continuum spectrum between the high and low
flux states was well described by a power-law of photon index
$\Gamma=2.0^{+0.2}_{-0.1}$.  This finding agreed with previous studies
that indicated that the spectral variability of MCG--6-30-15 in its
normal state is dominated by a power-law component (e.g. Fabian \etal
2002).  Second, and most importantly, ionized absorption models whose
continuum curvature mimics the red-wing of a broad iron line from $3-6
\keV$ were ruled out.  Such models generically predict strong
K$\alpha$ absorption lines of intermediately ionized iron.  Young
\etal showed that these lines are conclusively absent, falsifying the
ionized absorption model and further strengthening the relativistic
smearing hypothesis.

\section{The {\tt kerrdisk} Model}
\label{sec:newmodel}

\subsection{Model Description}
\label{sec:descrip}

As discussed in the introduction, we are motivated to construct an
iron line model that allows the spin parameter of the black hole to be
a free parameter and is maximally flexible.  

Our new relativistic emission line code, {\tt kerrdisk}, is written in
{\sc fortran77} so that it can be easily meshed with {\sc xspec}, and
has now been compiled successfully on both Solaris and Linux
platforms.  The dimensionless spin parameter of the black hole
($a=cJ/GM^2$; where $J$ is the angular momentum of a black hole of
mass $M$) can take on any value in the range $-1\leq a\leq 1$, where
negative values of $a$ correspond to a black hole that is rotating in
a retrograde sense relative to the accretion disk.  Although,
according to the equations of General Relativity, $a$ could have any
arbitrary value, the cosmic censorship hypothesis states that
naked singularities cannot exist in the universe, so a
black hole must be shrouded by an event horizon. This limits the
acceptable range of spin parameters to $-1 \leq a \leq 1$.  For
simplicity we consider only prograde spins up to the Thorne (1974)
spin-equilibrium limit, i.e., $0\leq a\leq 0.998$.

The limiting value of $a=0.998$ for black hole spins was first
discussed by Thorne (1974).  Therein, the author contends that if
black holes were simply accreting matter their spins would ascend to
$a\approx 1$ rather quickly, but because material in the accretion disk
radiates, and some of that emitted radiation is swallowed by the hole,
a counteracting torque is produced.  The origin of this counteracting
torque lies in the photon capture cross-section of the hole: black
holes have a higher capture cross-section for photons of negative
angular momentum (opposite to that of the hole itself) than for
photons of positive angular momentum.  Thus, the accretion of emitted
radiation from the disk results in an overall reduction in the angular
momentum of the hole until it reaches a theoretical equilibrium value
of $a=0.998$.  Recent work on magnetohydrodynamic (MHD) accretion
disks suggest that the continued transport of angular momentum from
matter within the radius of marginal stability, as well as angular
momentum lost from the rotating black hole itself via Blandford-Zjanek
like mechanisms (Blandford \& Znajek 1977), may lead to a rather
lower equilibrium spin (e.g., see the general relativistic MHD
simulations of Krolik, Hawley \& Hirose (2005)).  Equilibrium spins as
low as $a\sim 0.90$ are within the realm of possibility.

Following the method of Cunningham (1975), we compute the line profile
by employing a relativistic transfer function.  For a given black hole
spin $a$, disk inclination $i$, the observed accretion disk spectrum
can be written as
\begin{equation}
F_{\rm obs}(E_o)\propto
\int\frac{g^2}{\sqrt{g^*(1-g^*)}}I(E_o/g,\theta_e,r_e)\psi(r_e,
g^*;a,i)\,dg^*\,r_{\rm e} dr_e.
\end{equation}
Here, $I(E,\theta_e,r_e)$ is the rest-frame specific intensity of the
disk at radius $r_{\rm e}$ and energy $E$ emitted at a angle
$\theta_{\rm e}$ to the disk normal, $\psi$ is the Cunningham transfer
function which accounts for the effects of light-bending on the
observed solid angle from each part of the disk, $g$ is the ratio of
the observed to the emitted photon energy ($g=E_{\rm o}/E_{\rm em}$), and
${\it g^{*}}$ is the relative value of $g$ with respect to the
redshift extremes obtained from that annulus,
\begin{equation}
g^*=\frac{g-g_{\rm min}(r_{\rm e})}{g_{\rm
      max}(r_{\rm e})-g_{\rm min}(r_{\rm e})}.
\end{equation}
This formulation is attractive from the computational point of
view; light bending effects are isolated from beaming effects and
encoded in the transfer function $\psi(r_e, g^*;a,i)$ which is a
slowly varying function of its variables.  This allows us to perform
full computations of the transfer function at a relatively sparse set
of points in parameter space, and then use interpolation to accurately
determine the transfer function at a general point.

Specializing to the case of a $\delta$-function emission line allows
us to eliminate the $g^*$ integration from Eqn.~(1).  Assuming that the
surface emissivity of the emission line as a function of emission
radius $r_e$ and angle $\theta_e$ is $f(r_e,\theta_e)$, this gives
\begin{equation}
F_{obs}(E_o) \propto
\, \int\limits_{r_{\rm min}}^{r_{\rm max}}
\, \frac{g^{3}}
{\sqrt{g^{*}(1-g^{*}})}
\frac{1}{g_{\rm max}-g_{\rm min}} \,
\psi(r_e,g^*;a,i) \, f(r_e,\theta_e) 
\, r_{\rm e} dr_e.
\end{equation}

We use the algorithms of Speith \etal (1995) to compute the transfer
function $\psi(r_e, g^*; a, i)$ as well as $g_{\rm min}(r_e)$ and
$g_{\rm max}(r_e)$, thereby allowing us to perform this integration.
For a given spin $a$ and inclination $i$, high quality line profiles
(and all of the line profiles presented in this paper) are produced
using a transfer function computed at 50 radial bins, equally spaced
in the variable $1/\sqrt{r_{\rm e}}$, and 20 linearly spaced relative
redshift (${\rm g^*}$) bins across the line profile at any given
radius.  

These values were determined via rigorous trial and error to make the
best, smoothest possible line profile while keeping computation time
to a reasonable length.  Increasing either the number of radial or
relative redshift bins did not significantly improve the integrity of
the line profiles produced.  Also, because the transfer function varies
quite slowly with radius and relative redshift, greater frequency of
sampling was not warranted.

Our choice for the radial spacing corresponds to equal spacing of
non-relativistic Keplerian velocities.  We then evaluate the line
profile integral (Eqn.~3) with a densely spaced logarithmic grid in
$r_{\rm e}$, using linear interpolation in $r_{\rm e}$ of the (rather
sparsely sampled but slowly varying) transfer function.  In practice,
we use $5n_{\rm e}$-zones per decade of $r_{\rm e}$ where $n_{\rm e}$
is the number of frequency elements in the output line profile
(specified by the response matrix of the data being fit).
Experimentation shows that this produces high quality line profiles
even at very high resolution (i.e., large $n_{\rm e}$).

The Cunningham transfer function $\psi(r_e,g^*; a, i)$ is computed for
a $20\times20$ grid in $(a,i)$-space and is stored in a 40 MB table
that is accessed by the driver script for {\tt kerrdisk}.  Again, due
to the slowly varying nature of the transfer function, simple linear
interpolation to arbitrary spins and inclination angles produces
accurate and high quality line profiles.

For all of the models explored in this paper, we follow Fabian \etal 
(2002) and assume a line emissivity characterized by a broken
power-law between some inner radius $r_{\rm min}$ and outer radius
$r_{\rm max}$, i.e., 
\begin{eqnarray}
f(r_e)=\left\{
\begin{array}{ll}
0 & r_e<r_{\rm min}\\
(r_{\rm e}/r_{\rm br})^{-\alpha_1} & r_{\rm min}\le r_{\rm e}< r_{\rm br}\\
(r_{\rm e}/r_{\rm br})^{-\alpha_2} & r_{\rm br}\le r_{\rm e}< r_{\rm max}\\
0 & r_{\rm e}\ge r_{\rm max}
\end{array}
\right.
\end{eqnarray}

We note that, at present, the publicly available Speith algorithms on
which this work is based do not support the proper computation of
emission from within the radius of marginal stability, i.e., we are
restricted to $r_{\rm min} \ge r_{\rm ms}$.  Hence, all {\tt kerrdisk}
line profiles currently assume that the emission profile is truncated
within $r=r_{\rm ms}$.

The code takes $\sim 4 \s$ to produce a single line profile on a $2
\GHz$ processor linux machine.  When the model is used to fit data
within {\sc xspec}, several hundreds or thousands of iterations of the
code are often needed to get a good fit, resulting in run times on the
order of an hour or more.  In principle, this run time can be
significantly reduced by employing a parallelization scheme when
fitting.  Such a parallelization
scheme is possible to implement in the {\sc isis} spectral analysis
package (Houck 2002) using the PVM module (Michael Nowak and Andy
Young, private communication).

The best way to examine the results of the code line profile
simulations is to look at how the morphology of the line changes with
alterations made in certain important variables.  For illustrative
purposes, Fig.~1 shows model line profiles with a
rest-frame energy of $E = 1 \keV$.  The line profiles shown in this
figure have the same parameter settings as those shown in the review
by Reynolds \& Nowak (2003) to enable easy comparison.  The Reynolds
\& Nowak profiles are also based on the Speith algorithms but sample
the transfer function more sparsely and employ a cruder integration
technique to evaluate the line profile.  The fact that our line
profiles agree with, but are much less noisy than the Reynolds \&
Nowak profiles validates our newer integration/interpolation
technique.
\clearpage
\begin{figure}[h]
\centerline{
\includegraphics[width=0.4\textwidth,angle=90]{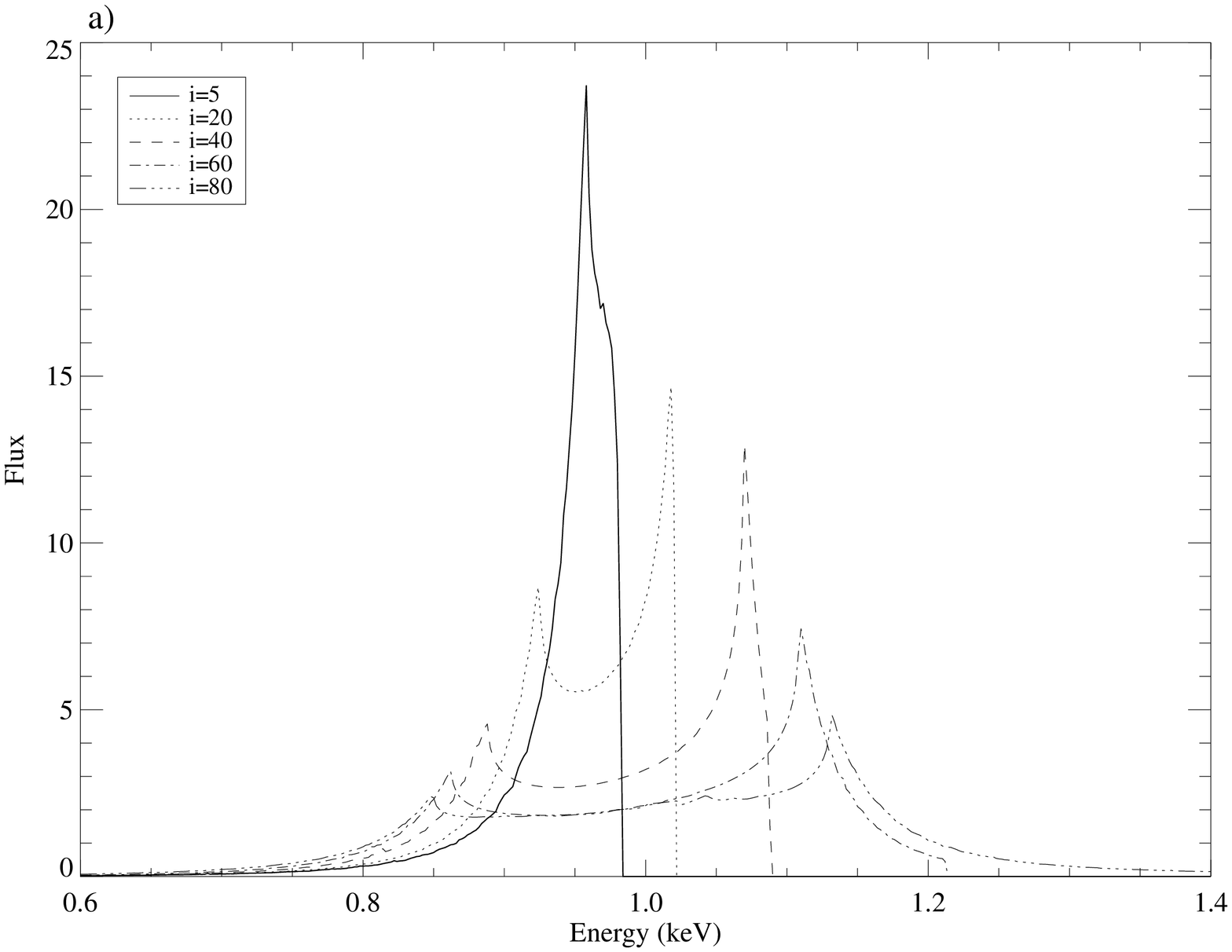}
}
\centerline{
\includegraphics[width=0.4\textwidth,angle=90]{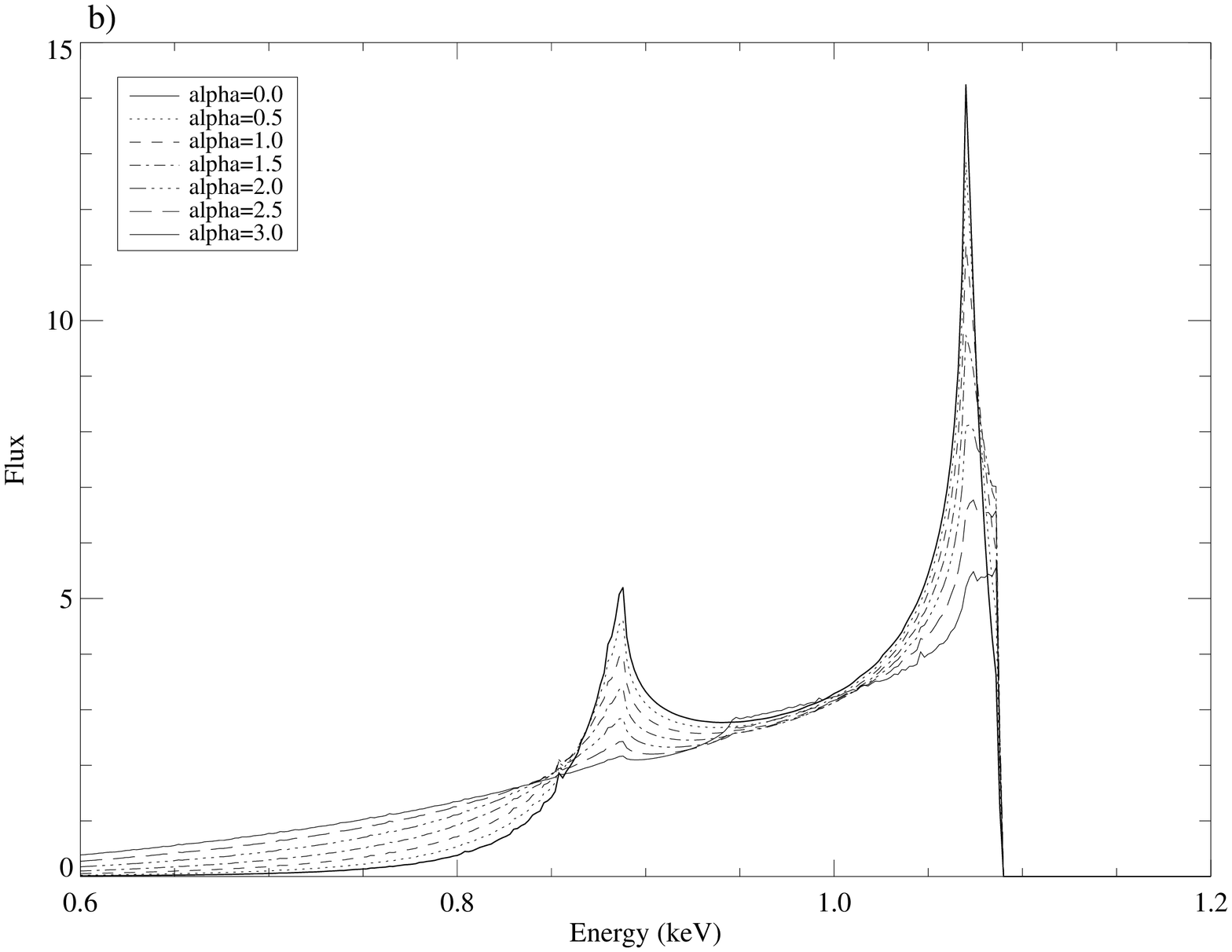}
}
\centerline{
\includegraphics[width=0.4\textwidth,angle=90]{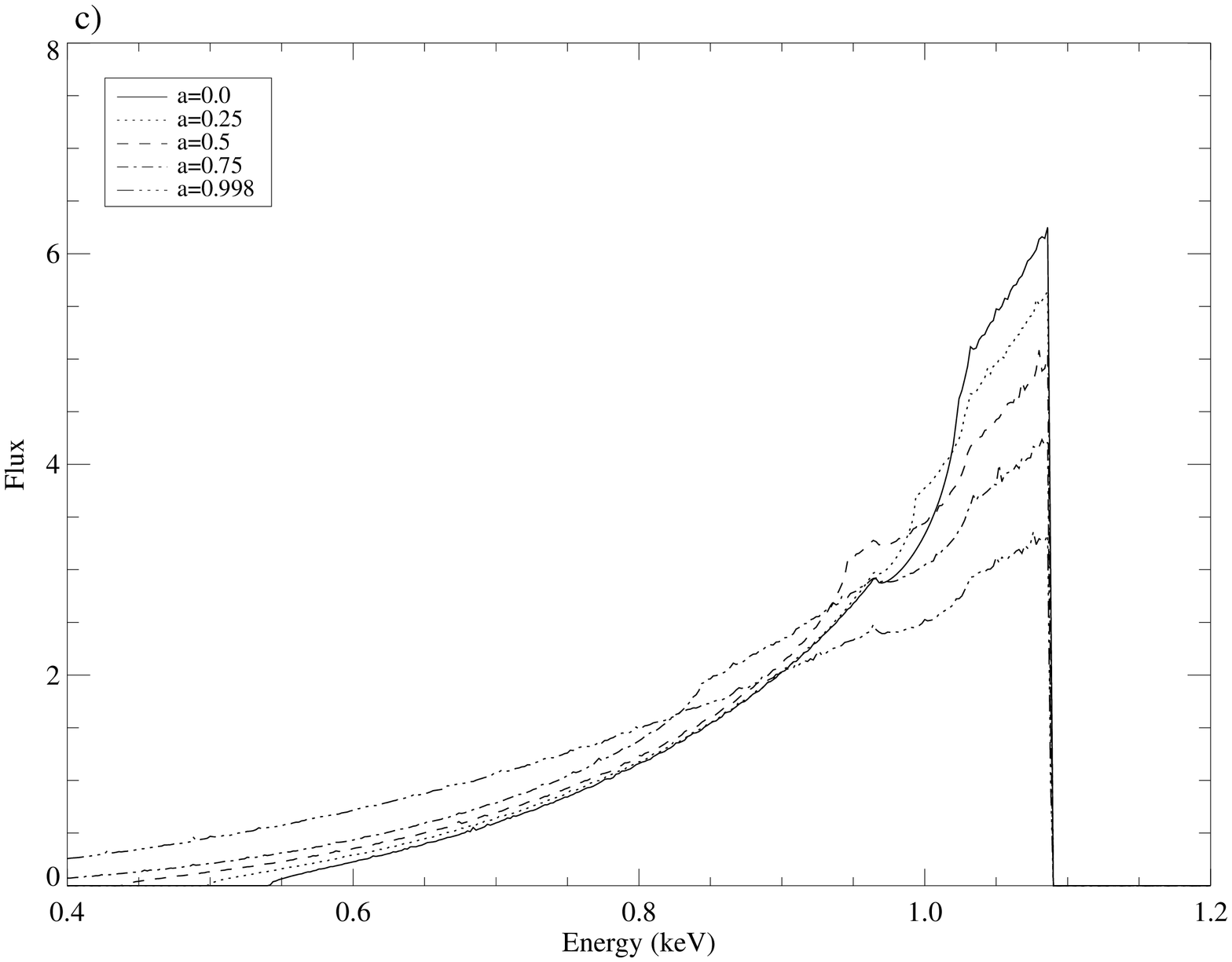}
}
\caption{Variation of the {\tt kerrdisk} line profile with (a) disk
  inclination angle, (b) disk emissivity index, and (c) black hole
  spin parameter.  In each case the range of emission in the disk is
  from $r = r_{\rm ms} - 50 \,r_{\rm g}.$ In (a), $a=0.998$, $\alpha = 1.5$.  In
  (b), $a = 0.5$, $i = 40$.  In (c), $i = 40$, $\alpha = 3$.}
\label{fig:kerr_params}
\end{figure}
\clearpage

\subsection{Comparison With Other Relativistic Disk Line Models}
\label{sec:improv}

Verification of our new line profile code can be demonstrated through
detailed comparisons with existing public models, such as {\tt laor} and {\tt
diskline} (where $a = 0.998$ and $0.0$, respectively), as well
as those of Dov\v{c}iak \etal (2004) and Beckwith \& Done (2004).  Since
the {\tt kyrline} model of Dov\v{c}iak has already been compared with the
{\tt kdline} model of Beckwith \& Done (2005), we need only to compare
{\tt kerrdisk} with {\tt kyrline}.  These comparisons are shown in
Figs.~2-5 and discussed in this Section.
\clearpage
\begin{figure}[h]
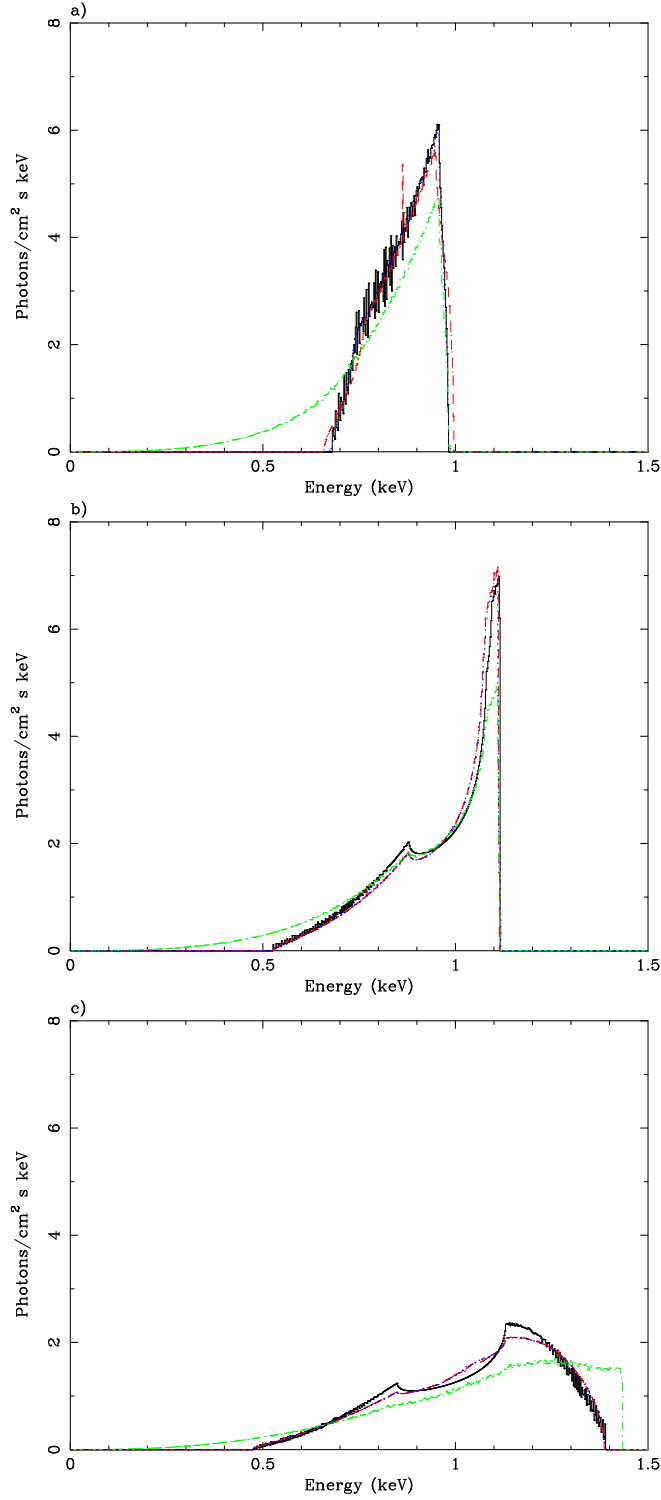

\centerline{
\includegraphics[width=0.4\textwidth,angle=270]{f2a.eps}
}
\centerline{
\includegraphics[width=0.4\textwidth,angle=270]{f2b.eps}
}
\centerline{
\includegraphics[width=0.4\textwidth,angle=270]{f2c.eps}
}
\caption{Various iron line models for a Schwarzchild black hole.  The
inclination angles represented are $5 \degmark$ for (a), $45 \degmark$
for (b), and $80 \degmark$ for (c).  Here $\alpha_1 = \alpha_2 = 3.0$,
and $r_{\rm min}$ and $r_{\rm max}$ are held constant at $6 \,r_{\rm
g}$ and $50 \,r_{\rm g}$, respectively.  The {\tt diskline} profile is
in solid black, the {\tt kerrdisk} profile is in dashed red, the {\tt
kyrline} profile including emission from within the ISCO is in
dash-dotted green and the {\tt kyrline} profile not including ISCO
radiation is in dotted blue.}
\label{fig:iplot_kvd}
\end{figure}

\begin{figure}[h]
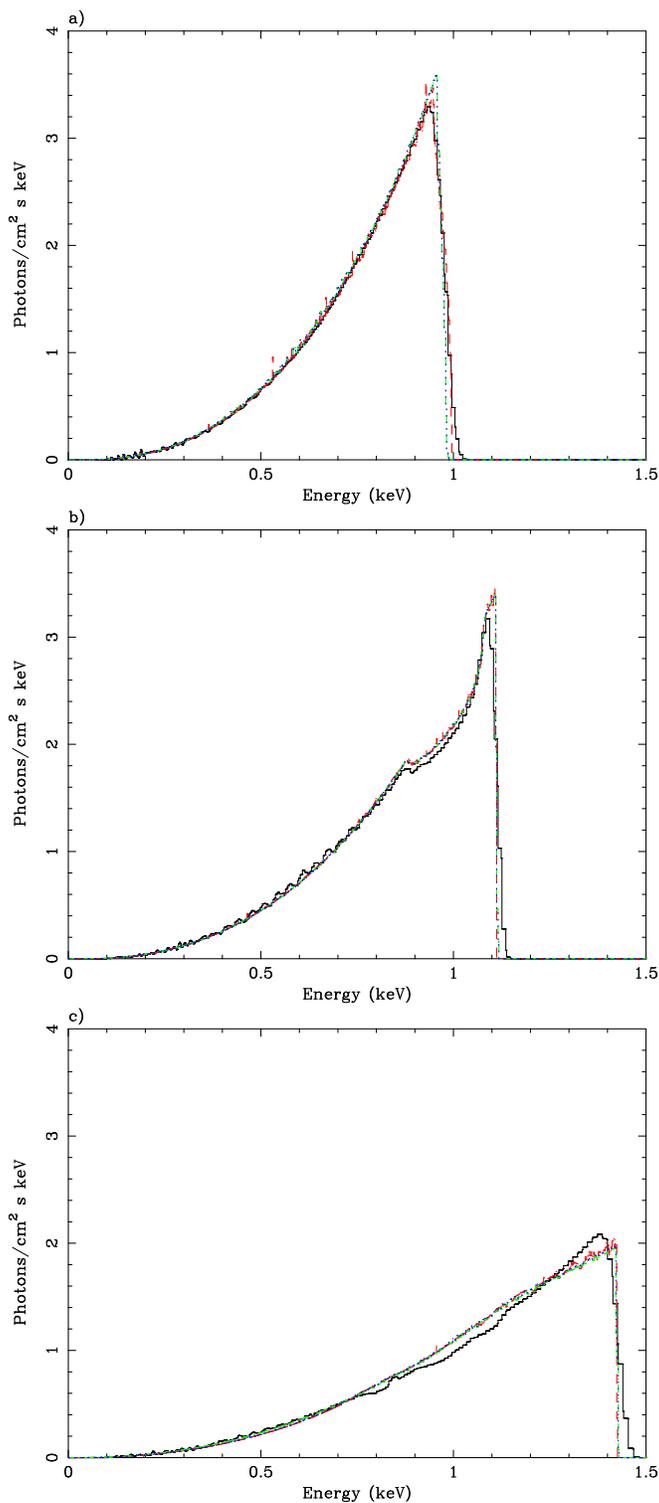

\centerline{
\includegraphics[width=0.4\textwidth,angle=270]{f3a.eps}
}
\centerline{
\includegraphics[width=0.4\textwidth,angle=270]{f3b.eps}
}
\centerline{
\includegraphics[width=0.4\textwidth,angle=270]{f3c.eps}
}
\caption{Various iron line models for a maximally spinning Kerr black
hole.  The inclination angles vary as above for the Schwarzchild case
in (a)-(c).  Other parameters are the same as those used in the
Schwarzchild case, but now $r_{\rm min} = 1.235 \,r_{\rm g}$,
corresponding to the radius of marginal stability for a maximal Kerr
black hole, rather than the $6 \,r_{\rm g}$ Schwarzchild $r_{\rm ms}$.
The {\tt laor} profile is in solid black and the rest of the color
scheme is the same as that used in Fig.~2 above.}
\label{fig:iplot_kvl}
\end{figure}
\clearpage
Given the lack of relativistic light bending in the {\tt diskline}
model, we expect our (fully relativistic) {\tt kerrdisk} line profiles
with $a=0$ to differ slightly from those computed with {\tt diskline},
especially at large inclination angles.  We indeed see slight
differences (Fig.~2).  We should stress, however, that the differences
are minor and {\tt diskline} should still be considered a perfectly
acceptable model for disks around Schwarzschild black holes for all
but the highest signal-to-noise data.  Examining the comparison of the
{\tt laor} models with the $a=0.998$ {\tt kerrdisk} model, we note
good agreement (Fig.~3).  The slight differences that do exist are
caused by an artificial smoothing of the {\tt laor} line due to
interpolation of a sparsely sampled transfer function.  Again,
however, we note that the {\tt laor} model is perfectly acceptable
model for disks around $a=0.998$ black holes for all but the highest
signal-to-noise data.

The real power of this new generation of line profile models is the
freedom in the spin parameter of the black hole, therefore the real
verification of our code lies in a detailed comparison of {\tt
kerrdisk} and {\tt kyrline}.  Indeed, as shown in
Figs.~2-3, the
line profiles for {\tt kerrdisk} and {\tt kyrline} are virtually
indistinguishable when one does not include radiation from within the
radius of marginal stability in {\tt kyrline}.  Also shown in these
figures are the effects of including emission from within the radius
of marginal stability (down to the horizon, for the sake of
illustration), as computed by Dov\v{c}iak \etal (2004).  
This has the greatest effect on the line profiles from
the slowly spinning holes since it is only in these systems that an
appreciable fraction of the plunging
region is subject to modest (rather than
extremely large) gravitational redshift.  We intend to extend our
model to include the region within the radius of marginal stability in
future work, carefully considering the effects of both the level of
ionization and the optical depth of the material within the plunging
region, in particular, as discussed in \S\ref{sec:intro}.  
Both of these effects can have a
substantial impact on the contribution of emission from this region to
the overall line profile.

\subsection{The Convolution Model}
\label{sec:conv}
 
In reality, the irradiation of the disk by the primary X-ray source
results in a whole X-ray reflection spectrum consisting of Compton and
radiative recombination continua plus numerous fluorescent and
radiative recombination lines.  The iron line is the most prominent
due to its high rest-frame energy and its intrinsic strength, but many
other species are also excited (e.g. oxygen, nitrogen, silicon and
sulfur).  The whole X-ray reflection spectrum from the disk will be
subject to the same extreme Doppler and relativistic processes that
combine to alter the morphology of the iron line.  The most physically
realistic simulation of the reflected spectrum from a photoionized
disk surface has been presented by Ross \& Fabian (2005) --- these
high-quality models capture the ``traditional'' X-ray reflection
processes (Compton scattering, photoelectric absorption and
fluorescent line emission) as well as the powerful soft X-ray
radiative-recombination line emission expected from an X-ray
irradiated photoionized surface of an optically-thick accretion disk.
Ross \& Fabian (2005) have provided their results in the form of
tabulated spectra that can be used in {\sc xspec}.  We then wish to
convolve this spectrum in velocity space with a relativistic smearing
kernel such as the one used to generate the {\tt kerrdisk} line
profile.  To facilitate this, we have also produced a convolution form
of our line profile model, {\tt kerrconv}, whose results mirror
those of the {\tt kyconv} model designed by Dov\v{c}iak \etal (2004)
and the {\tt kdconv} model of Beckwith \& Done (2004), provided that
no radiation from within the ISCO is included in the latter two
models, as was the case for the comparison of {\tt kerrdisk} with {\tt
kyrline} and {\tt kdline} in \S\ref{sec:improv} above.  As with {\tt
kerrdisk}, the {\tt kerrconv} parameters include emissivity
indices for the inner and outer disk separated by a break radius,
inner and outer radii for the disk emission, the spin parameter of the
black hole, and the inclination angle of the disk with respect to our
line of sight.  The {\tt kerrconv} model is readily validated by
applying it to a narrow emission line and comparing the resulting
spectrum with the regular {\tt kerrdisk} model.

\section{Determining the Spin of the Black Hole in MCG--6-30-15}
\label{sec:MCG6}

In this Section, we use our new models ({\tt kerrdisk} and {\tt
kerrconv}) along with the model of Dov\v{c}iak \etal (2004) to
confront the issue of determining the spin of the black hole in
MCG--6-30-15.  Because of the extremely robust, broad, well-studied
iron line in this system, MCG--6-30-15 is an excellent candidate for
such a study.  Given the complexity of the spectrum displayed by the
source, it is important to perform this exercise in a step-by-step
manner, clearly enumerating all of the assumptions at each stage, and
employing physical models for the spectral complexity whenever
possible.

This guides the study presented in this Section.  Due to the
unprecedented signal-to-noise, we use the EPIC-pn data from the
aforementioned long {\it XMM-Newton} observation of MCG--6-30-15.
Data preparation and reduction followed Vaughan \& Fabian (2004)
exactly.  In this Section, we present a step-by-step analysis of these
data using models of increasing complexity and physical realism.
Initially, to illustrate the potential power of broad iron lines for
spin determination, we modeled the $2-10 \keV$ EPIC-pn as a simple
power-law continuum modified by a broad iron line (and absorption by
the Galactic column of $4.1 \times 10^{20} \pcmsq$ toward this
source); this is comparable to the study performed by Dov\v{c}iak
\etal (2004).  It does, however, neglect the significant effects that
continuum curvature due to ionized absorption could have on the
inferred iron line parameters (and hence inferred black hole spin).
To assess these effects, we next model the effects of multiple warm
absorbers and dust on first the $2-10 \keV$ spectrum and then the full
$0.6-10 \keV$ spectrum.  In our most sophisticated spectral model, we
describe the $0.6-10 \keV$ band including multiple absorption
components and augmenting the simple broadened iron line with a
relativistically smeared ionized X-ray reflection spectrum from Ross
\& Fabian (2005).

The best-fit parameters and error bars for each progressive model fit
are shown in Table~1.  Each fit and the corresponding $\Delta \chi^2$
with respect to changes in $a$ in each case are shown in Figs.~6-10.
The unfolded spectrum and the best-fit model components for both the
single {\tt kerrdisk} case and the full ionized reflection spectrum
convolved with {\tt kyconv} are shown in Figs.~11-12.  For the warm
absorber tables described in Table~1 and \S4.2, all abundances are
frozen at the solar value.  For all model components, the redshift is
set to $z = 0.008$, the optically determined value for MCG--6-30-15
(Reynolds et al. 1997).  In all of the fitting below, the inner disk
radius contributing to the iron line emission (or X-ray reflection
spectrum in the case of Model~5) was not allowed to be smaller than
the radius of marginal stability.

\subsection{Simple Power-Law Continuum and Iron Line Across the $2-10
  \keV$ Spectrum}
\label{sec:simplepo}

Initially, we perform an analysis of the $2-10 \keV$ spectrum assuming
that the underlying continuum is a simple power-law (absorbed by the
Galactic hydrogen column) and that the disk spectrum is just a single
iron line (rather than a whole reflection spectrum).  Fig.~4 shows the
hard spectrum as fit by this photoabsorbed power-law (Model~1).  To
accurately model the continuum we have initially ignored the $4-7 \keV$
range when fitting this component.  This prevents any contamination of
the fit by the presence of an iron line reflection signature.  Once
the fit was complete, energies from $4-7 \keV$ were included again.
As suspected, the data/model ratio shown in the lower panel demonstrates a
significant residual feature above the continuum, which appears to have 
the form of a highly
broadened iron line peaking at $6.4 \keV$.  The presence of this large
residual feature results in a poor fit, $\chi^2/{\rm dof}=4577/1106$.
Fig.~5a plots the data/model ratio again, this time for a model
including a single, broad {\tt kerrdisk} line with a rest-frame energy
of $6.4 \keV$ (Model~2).  Two narrow redshifted gaussians representing
a cold Fe-K$\alpha$ line and an ionized line of iron are also included
at $6.4$ and $6.9 \keV$, respectively, as in Fabian \etal (2002).  The
$6.4$ and $6.9 \keV$ lines both have equivalent widths of $\sim 14
\eV$.  

It should be noted here that Fabian \etal (2002) acknowledged that a
$6.74 \keV$ absorption line of $EW=-138 \pm 35 \eV$ fit the data as
well as an emission line at $6.9 \keV$ and $EW=18 \pm 6 \eV$.  This
absorption feature was also preferentially used in the MCG--6-30-15
work of Vaughan \& Fabian (2004), and was consistent with the
prediction of Sako \etal (2003) based on an RGS observation of this
source in 2001.  However, the $6.7 \keV$ absorption line detected in the
deep high-resolution {\it Chandra}/HETGS spectrum of Young \etal
(2005) is significantly weaker than that fitted by Fabian \etal
(2002), with equivalent widths of $EW=-18^{+7}_{-5}, -13 \pm 9, and
-25 \pm 9 \eV$ during the average, low and high flux states of the
source, respectively.  Thus, the high-resolution {\it Chandra}
spectrum does {\it not} support a spectral model for the EPIC spectrum
in which the complexity in the $6.6-7.0 \keV$ range includes a very
strong absorption line.  Two possible loopholes in this argument are
(1) an order of magnitude temporal change in the helium-like iron
column density between the {\it XMM-Newton} and {\it Chandra} grating
observations and (2) extreme $\sim 10^4 \kmps$ velocity broadening of
the absorption feature which would diminish its detectability in the
high-resolution spectrum.  More plausible is the notion of a weak
($EW \sim 40 \eV$) and slightly broad emission line from hydrogen like
iron (Fe\,{\sc xxvi}).  Given that {\it Chandra} shows there to be a weak
($EW \sim -20 \eV$) narrow Fe\,{\sc xxvi} absorption line at $6.97 \keV$, the
EPIC-pn spectrum would be expected to show a net emission feature with
$EW \sim 20 \eV$.

Most importantly, however, the details of whether this spectrum
complexity is described by an ionized iron emission or absorption line
has almost negligible effect on the broad iron line.  In terms of the
effect on the overall fit, Fabian \etal (2002) note that the
relativistic line parameters differ insignificantly when one employs
an absorption line rather than an emission line in the model.  We have
checked this result in our own analysis; specifically, we have
replaced the $6.9 \keV$ gaussian in Model~4 with a $6.74 \keV$
gaussian in absorption (i.e., negative flux).  The result was a
$\Delta\chi^2/\Delta{\rm dof}=+26/0$, indicating a marginal decrease
in the overall goodness of fit.  Visually, this fit was
indistinguishable from the best fit from Model~4, and as per Fabian
\etal, we also found minimal change in the relativistic line
parameters.  The inner emissivity index of the disk became marginally
steeper and its inclination angle increased very slightly, but the
changes were well within the statistical error bars.  The most
interesting point, however, is that the best fit equivalent width of
this absorption line was only $-21.3 \eV$; much less than the $-138
\eV$ found by Fabian \etal.  Such a modest equivalent width in
comparison to what should be necessary suggests that this line is not
robustly wanted in our fit to the data.

The three iron lines (two narrow and one broad) that we do choose to
include significantly improve the fit ($\chi^2/{\rm dof}=960/1096$),
and succeed in modeling out the residual feature shown in Fig.~4.  At
$90\%$ confidence for one interesting parameter, our best fit for
Model~2 indicates a very rapidly rotating black hole
($a=0.970^{+0.003}_{-0.015}$).  To gauge the sensitivity of this fit
to the spin of the black hole, Fig.~5b plots the change in the
goodness of fit parameter $\Delta\chi^2$ as a function of black hole
spin parameter.  This clearly demonstrates that $\chi^2$ improves
dramatically as one approaches very rapid spins.  The equivalent width
of the broad iron line in our best-fitting model is $\sim 729 \eV$,
quite a bit higher than the $\sim 550 \eV$ cited by Fabian \etal
(2002), the $450 \eV$ found by Ballantyne \etal (2003), or the $\sim
250 \eV$ found by Dov\v{c}iak \etal (2004).  This unphysically high
equivalent width is clearly due to the simplicity of this model ---
the effects of absorption and the X-ray reflection continuum, in
particular, will introduce additional curvature into the continuum,
thereby alleviating the need for such a strong line.

\subsection{Modeling the Warm Absorber}
\label{sec:WA}

In order to accurately assess the width and morphology of the iron
line in an AGN, it is imperative that the soft portion of the spectrum
be modeled correctly.  One must be concerned about confusing a broad
red-wing of a relativistic iron line with continuum curvature
resulting from a putative ``warm absorber'' present within the AGN
system.  In fact, it has been argued that broad iron lines may be
entirely unnecessary once one correctly accounts for the effects of a
warm absorber (see Sako \etal 2003 for a discussion of MCG--6-30-15;
see Turner \etal 2005 for a discussion of NGC 3516).  As mentioned in
\S\ref{sec:history}, a deep {\it Chandra}/HETGS observation of
MCG--6-30-15 fails to find the K-shell absorption lines of the
intermediate charge states of iron predicted from a model in which a
warm absorber is mimicking the whole relativistic red-wing of the iron
line.  However, the question remains as to the effects that warm
absorption has on fitting of the extreme red-wing of the iron line
which drives black hole spin constraints.  Very long data sets are
needed in order to obtain spectra with enough signal to put both broad
line and warm absorber models to the test and pursue the question of
their overlap with statistical validity.  The $\sim 350 \ks$
observation of MCG--6-30-15 taken by Fabian \etal (2002), which we use
here, is ideal for such a study because of its large number of counts
and its resolution of the broad iron feature this galaxy is thought to
possess.

We have used the {\sc xstar} spectral synthesis package for
photoionized gases (version 2.1kn3, Kallman 2005) to construct a grid
of warm absorber models as a function of the absorbed column density
$N_{\rm H}$ and ionization parameter $\xi$.  The ionization parameter
is given by the usual definition,
\begin{equation}
\xi=\frac{L_{\rm i}}{n_{\rm e}r^2},
\end{equation}
where $L_{\rm i}$ is the luminosity above the hydrogen Lyman limit,
$n_{\rm e}$ is the electron number density of the plasma and $r$ is the
distance from the (point) source of ionizing luminosity.  We constructed
$20\times 20$ grids of models uniformly sampling the $(\log\,N_{\rm
H},\log\xi)$ plane in the range $N_{\rm H}=10^{20}\rightarrow
10^{24}\pcmsq$ and $\xi=1\rightarrow 10^4 \ergpcmps$.  While these
were made to be multiplicative models (i.e., absorber models that can
be applied to any emission spectrum), the ionization balance was
solved assuming a power-law ionizing spectrum with a photon index of
$\Gamma=2$.  This is a good approximation to MCG--6-30-15.

Initially, we side-step the complexities of the soft ($<2 \keV$)
spectrum and apply this warm absorber model to the $2-10 \keV$ only.
In many ways, neglecting any constraints from the spectrum below $2
\keV$ maximizes the impact that it may have on the broad iron line;
the sole ``job'' of the absorption component in this setting is to
attempt to fit the curvature of the spectrum otherwise attributed to
the broad iron line.  We do notice a modest reduction in the goodness
of fit parameter compared with Model~2 (the simple power-law and {\tt
kerrdisk} model) $\Delta\chi^2/\Delta{\rm dof} = -26/2$.  The warm
absorber in the best fitting Model~3 is of modest optical depth and
rather weakly ionized: $N_{\rm H}=4.22 \times 10^{22} \pcmsq$ and log
$\xi=0.84$.  Although the change in the goodness of fit is not
dramatic, the additional continuum curvature introduced by the warm
absorber leads to a reduction in the equivalent width of the iron line
from $EW=729 \eV$ down to the more physically reasonable value
$EW=521\eV$.  As can be seen from Table~1, the parameters that
determine the best-fitting shape of the iron line (the emissivity
indices, inner radius, break radius, outer radius, inclination of the
disk and black hole spin) are essentially unaffected by the inclusion
of this warm absorbing component.  As shown in Fig.~6, a rapidly
rotating black hole is still preferred in this case
($a=0.997^{+0.001}_{-0.035}$).  However, in contrast to the case with
Model~2, the inclusion of a warm absorber to the $2-10 \keV$ spectrum
allows models with low black hole spin to fit the data adequately
(with low-spin cases producing a goodness of fit parameter which is
only $\Delta\chi^2 \sim 40$ worse that high-spin cases) --- in these
cases, the $3-4 \keV$ curvature is being modeled as the effects of
warm absorption rather than the extreme red-wing of the iron line.

Of course, the soft X-ray band ($<2 \keV$) is extremely important for
constraining the properties of warm absorbers; the opacity of most
well-studied warm absorbers is dominated by oxygen and iron edge/line
absorption in this band.  Hence, to be complete, we must extend our
study of the effects of warm absorption on X-ray reflection features to
the full $0.6-10 \keV$ band.

When fitting the $0.6-10 \keV$ band, we find that we {\it cannot}
produce an acceptable fit with a model consisting of a simple
power-law and {\tt kerrdisk} line subjected to the effects of Galactic
absorption and a one-zone warm absorber.  This is not surprising: it
is generally thought that warm absorbers must be physically more
complex than a one-zone model can account for, i.e., they cannot be well
described by a single value of the column density and ionization
parameter.  Physically, the warm absorber likely represents an wind
emanating from the accretion disk and/or cold torus surrounding the
central engine and may well contain dust.  A continuum of ionization
parameters is likely to exist along the line of sight.  For
computational purposes, however, it is convenient to approximate this
as a discrete number of zones, each of which is characterized by a
column density and ionization parameter.  Adding in a second warm
absorber dramatically improves the fit and, indeed, the two absorbers
do seem to represent distinct ``zones'' of material based on their
column densities and ionization parameters (see Table~1 for details).  

Even taking both warm absorber models into account, however, there
still appears to be significant remaining absorption in the spectrum
below $2 \keV$, as well as a strong soft excess below $\sim 0.7 \keV$.
A strong edge due to the $L_3$-edge of neutral iron (presumably in
dust grains embedded within the warm absorber) has already been noted
in high-resolution {\it Chandra} and {\it XMM-Newton} grating spectra
of MCG--6-30-15 (Lee \etal 2002; Turner \etal 2003).  Incorporating this
edge into our fit (employing spectral tables kindly provided to us by
Julia Lee) makes a significant visual and statistical improvement, 
largely explaining the unmodeled absorption mentioned above.  It
is worth noting that the edge demands quite a high column density of
iron: $\log (N_{\rm Fe})= 17.54$, which is approximately a factor of
two higher than that found by Turner \etal (2003).  To address the
soft excess seen below $0.7 \keV$, we employ a simple blackbody model.
Since the data are only sensitive to the tail of this component, the
other parameters describing the spectrum are completely insensitive to
the precise model used for this soft excess.  Table~1 details the fit,
which includes two warm absorption zones, the Fe-$L_3$ edge and the
additional soft excess.  

Although we cannot statistically compare the Model~3 and Model~4 fits
(since we have expanded the energy range of study between these
models), Model~4 does appear to describe the full $0.6-10 \keV$
spectrum very well (see Fig.~7).  As before, the data strongly prefer
a rapidly rotating black hole.  The inclusion of the $0.6-2 \keV$ data
apply extra constraints on the warm absorbers; the partial degeneracy
found in Model~3 between the red-wing of the iron line and the
curvature introduced by warm absorption is now removed.  At $90\%$
confidence, this model gives formal constraints on the black hole spin
of $a=0.997 \pm 0.001$, and the broad iron line has an
equivalent width of $\sim 926 \eV$.  This is significantly broader
than we find for Models~2-3, or in any of the other analyses of this
data set, and reflects the breadth of the red-wing of the iron line
feature in this fit.  The values for the best-fit parameters for
Model~4 are shown in Table~1.

Fig.~9 shows the unfolded spectrum for MCG--6-30-15 fit with Model~4
using a simple {\tt kerrdisk} line.  Each model component is colored
and labeled separately to highlight its relative contribution to the
fit.  Note the relatively strong blackbody component that must be included at
$\sim 0.1 \keV$ in order to accurately model the spectrum below $\sim
2 \keV$, as well as the redshifted gaussian emission lines at $6.4$
and $6.9 \keV$ that must be added to the broad neutral iron line to
fully capture the shape of the hard spectrum.

Turner \etal (2003) have approached the question of absoprtion in
MCG--6-30-15 by analyzing the RGS spectrum from the same {\it
XMM-Newton} observation used here (from Fabian \etal, 2002).  In
fitting the $0.32-1.7 \keV$ range, these authors have identified six
components of absorption: absorption by the cold Galactic column, four
``zones'' of warm absorbing plasma, and an absorption L$_3$ edge of
neutral iron (from dust embedded in one or more of the warm absorbing
zones).  Considering this detail in structure identified by Turner
\etal, it might appear that our spectral model (which only requires
two warm absorber zones plus the neutral iron-L$_3$ edge to describe
the spectrum) is inconsistent with the picture painted by the RGS.
Upon closer inspection, however, we see that this is not the case.
Firstly, the lowest ionization warm absorber seen in the RGS
($\log \xi \sim -4.42$) cannot be distinguished from neutral absorption
by our $0.6-10 \keV$ EPIC-pn spectrum and hence is, in fact, accounted
for through the neutral absorption column present in our model.
Secondly, the two highest ionization warm absorbers identified by the
RGS actually have rather similar ionization states ($\log \xi \approx
1.6-1.7$ for model-2 of Turner \etal, 2003) and are only separated
into two zones through their kinematics; they are separated by $\sim
2000 \kmps$ in velocity space by the RGS.  The EPIC-pn instrument,
however, would not be able to resolve the velocity difference of these
two zones.  Accounting for these two facts, we would expect the Turner
\etal four-zone RGS model to reduce to a two-zone model when applied
to EPIC-pn data.

The column densities and ionization parameters are lower in the Turner
\etal fit than in ours, but unfortunately it is difficult to compare
these values in a meaningful way due to calibration issues in the
continuum response that presently exist with RGS data.  This renders
it nearly impossible to perform simultaneous RGS/EPIC-pn fits in {\sc
xspec}, which is why we have chosen not to address such a joint fit
within the scope of this paper.  Cross-calibration issues similarly
affect our ability to compare EPIC-pn data with {\it Chandra}/HETGS
results, making it difficult to obtain a more precise, independent
check on the fit to energies below $\sim 2 \keV$.

\subsection{Model Including a Full Reflection Spectrum}
\label{sec:ion_disk}

The broadened iron emission line of Models~1-4 is, of course, just
the tip of the iceberg; the disk produces a whole spectrum of
fluorescent and recombination lines, radiation-recombination continua
and Compton backscattered continuum.  To obtain truly reliable
constraints, we must consider the full X-ray reflection spectrum.
 
In Model~5, we take the basic continuum/absorption components of
Model~4 and augment the simple iron line with the full X-ray
reflection spectrum from an ionized disk surface (Ross \& Fabian
2005).  The Ross \& Fabian (2005) models describe the reflected
spectrum emitted by an optically-thick atmosphere (here, the surface
of an accretion disk) of constant density that is illuminated by
radiation with a power-law spectrum (here, photons that have been
inverse Compton-scattered by relativistic electrons in the corona or
base of a jet).  We then convolve this reflection spectrum with the
effects of relativistic smearing via {\tt kerrconv}.
Interestingly, as will be noted below, the soft X-ray emission
associated with the photoionized disk surface naturally explains the
soft excess without the need for an additional ad-hoc blackbody
component.  Hence, Model~5 does not include the blackbody component
of Model~4.

The best-fit parameters for this model are shown in Table~1, and in
comparison with Models~1-4 it appears that Model~5 provides a
statistical fit to the data that is not as good: ($\chi^2/{\rm
dof}=1793/1374$; $\Delta\chi^2=+51$ for one more degree of freedom
compared with Model~4).  Model~5 is, however, our most physical model
in the sense that the whole reflection spectrum is treated as opposed
to just the iron line, resulting in a natural explanation for the soft
excess.  That is to say that both the somewhat arbitrary blackbody
component and the broad {\tt kerrdisk} line of Model~4 are not
required in this fit; the soft excess and broad iron feature are
instead both fully described by the smeared radiative recombination
line/continuum emission from the irradiated accretion disk.  This
change in the modeling of the soft excess increases the photon index
of the continuum power-law component to $\Gamma=2.09$, and also
results in an increase in the inferred depth of the neutral iron-$L_3$
edge to an iron column density of $\log(N_{\rm Fe})=17.68$ (over
$17.54$ in Model~4).  The column densities and ionization parameters
of the two included warm absorbers also vary somewhat from Model~4,
but the clear delineation between them remains evident.  See Table~1
for details.
 
The inclusion of an ionized X-ray reflection spectrum also has
important implications for the derived spin parameter, as can be seen
in Fig.~10.  The fact that a significant component of the line
broadening in this model is now due to Compton scattering reduces the 
inferred black hole spin to $a=0.989^{+0.009}_{-0.002}$, slightly lower than 
the value determined in Model~4, but still consistent with a very 
rapidly spinning black hole.  We do find, again, that a narrow
Fe-K$\alpha$ line at $6.4 \keV$ is necessary in order to properly
model the shape of the spectrum, as well as a $6.9 \keV$ line of
ionized iron as in Model~4.  The equivalent widths of these narrow
lines are $24.7 \eV$ and $27.3 \eV$, respectively.  Including all of
these components, we find that the total $0.6-10 \keV$ luminosity of
MCG--6-30-15 is $L_{\rm X}=9.34 \times 10^{42} \ergps$ using WMAP
cosmological parameters.

Fig.~10 shows the plot of the relative contributions of the best-fit
model components for Model~5.  The main features are an ionized disk
reflection spectrum relativistically blurred with a {\tt kerrconv}
convolution model to represent the soft emission and ionized iron
features, as well as two {\tt zgauss} components to model the cold,
neutral iron line at $6.4 \keV$ and the $6.9 \keV$ line included in previous
fits after Fabian \etal (2002).  The absorption components in the soft
spectrum are the same as those used for Model~4.  Note that the
presence of the ionized disk reflection negates the need for the
blackbody component of Model~4 shown in Fig.~9.  

While not essential for the principal issue of this paper (i.e.,
determining black hole spin), it is instructive to estimate the
``reflection fraction'' of the ionized reflection spectrum, $R_{\rm
refl}$. This parameter is defined to be proportional to the ratio of
the normalization of the reflection spectrum to that of the intrinsic
spectrum, and normalized such that $R_{\rm refl}=1$ corresponds to a
reflector that subtends half of the sky as seen from the X-ray source.
Operationally, the ionized reflection model of Ross \& Fabian (2005)
is characterized by an absolute normalization and hence one cannot
fit trivially for $R_{\rm refl}$.  We estimate this parameter by
extending Model~5 out to $100 \keV$ and setting $R_{\rm refl}$ to be the
ratio of the normalization of the reflected and intrinsic spectra at
the peak of the Compton reflection hump at $35 \keV$.  This technique
yields $R_{\rm refl} \approx 1.25$.  Previous studies have also found
the reflection parameter to be in this range for MCG--6-30-15, but it
should be noted that in these cases $R_{\rm refl}=\Omega/2\pi$ has
been a fitted parameter in the reflection model used, and as such has
been considered an indication of the covering fraction of the
reflecting material.  For example, Fabian \etal (2002) began their
spectral fitting by assuming $R_{\rm refl}=1$, but discovered that
when this parameter was left free they achieved better statistical
fits to the data.  The best-fit value determined by these authors was
$R_{\rm refl}=2.2^{+1.1}_{-0.7}$, which is consistent with our own,
within error bars.  By contrast, Lee \etal (2000) used a $\sim 400 \ks$
simultaneous {\it ASCA}/{\it RXTE} observation of MCG--6-30-15 to
identify four distinct spectral states for the source.  These authors
noted that as the flux increased, $\Gamma_{3-10}$ steepened while
$\Gamma_{10-20}$ gradually flattened at approximately the same rate.
Lee \etal modeled the reflection with a {\tt pexrav} component in
{\sc xspec}, using a reflection inclination angle matching that of the
accretion disk at $i=30 \degmark$ and possessing an exponential cutoff
for the Comptonizing source power-law at $100 \keV$.  The iron
abundance was maintained at twice the solar value.  These authors
found that as the source flux increased, $R_{\rm refl}$ did as well,
beginning at $0.35^{+0.15}_{-0.16}$ and topping off at
$1.37^{+0.23}_{-0.14}$.  The last two flux states are consistent with
our result within error bars.  

A very high iron abundance in this disk is required by our Model~5 fit
($Z_{\rm Fe}>9.8\,Z_\odot$ at $90\%$ confidence).  Previous studies
have also detected an overabundance of iron in the disk of
MCG--6-30-15 (e.g. Ballantyne \etal 2003), though none have proposed
this high an abundance.  If we freeze the iron abundance to be
$3\,Z_\odot$ and re-fit, as in Ballantyne \etal (2003), the goodness
of fit parameter is increased by $\Delta\chi^2=+463$ for one less
degree of freedom, and there are
obvious residual features created in the model fit by employing this
tactic.  Most of the absorption parameters remain relatively
unchanged, but the disk parameters are altered considerably: $r_{\rm
  min}=1.88\,r_{\rm g}$ and $r_{\rm max}=12.26\,r_{\rm g}$, so the
disk only radiates over a fairly thin ring near the radius of marginal
stability.  The emission profile consists of $\alpha_1=5.60$,
$\alpha_2=1.00$, and $r_{\rm br}=3.22\,r_{\rm g}$, reinforcing this
conclusion.  The inclination angle of the disk is reduced to $\sim 20
\degmark$ from $\sim 30 \degmark$ in the best-fitting Model~5, and the
spin of the black hole is considerably lowered to $a=0.31$.  Even
though this is the best fit with $Z_{\rm
  Fe}=3\,Z_\odot$, it is clear from the form of the residuals around $6.4 \keV$
that the reflection spectrum is being insufficiently broadened.
Based on this fact and the substantial worsening of $\chi^2$, it appears that
our fit indeed prefers a higher abundance of iron than that found by
Ballantyne \etal.

Our confidence in the goodness of fit provided by Model~5 is
strengthened by the consistency with which the continuum model we have
employed extends to fit the {\it BeppoSAX}/PDS spectrum of
MCG--6-30-15 at high energies.  The {\it BeppoSAX} observation was
taken simultaneously with the {\it XMM-Newton} in 2001 and published
by Fabian \etal (2002), acting as a valued ``sanity check'' to the
derived EPIC-pn fit by providing spectral data from $\sim 15-100
\keV$.  We have performed a joint fit to the PDS and pn data within
{\sc xspec}, applying our Model~5 to both sets of data.  Even with the
paucity of counts at high energies, the results of this fit clearly
show that Model~5 provides an excellent description of both the pn and the
PDS continuum.  The ability to place such additional constraints on
our continuum and absorption parameters gives us greater confidence in
the constraints we have correspondingly derived for the broad iron
line parameters, including black hole spin.
\clearpage
\thispagestyle{empty}
\setlength{\voffset}{35mm}
{\rotate
\begin{table}
\begin{center}
\caption{Best-fit parameters for MCG--6-30-15}
\begin{scriptsize}
\begin{tabular} {|l|l|l|l|l|l|l|} 
\hline \hline
{\bf Model Component} & {\bf Parameter} & {\bf Model~1} & {\bf
  Model~2} & {\bf Model~3} & {\bf Model~4} & {\bf Model~5} \\
\hline \hline
{\tt phabs}    & $N_{\rm H} (\pcmsq)$ & $0.041$ & $0.041$ & $0.041$ &
$0.041$ & $0.041$ \\
\hline
{\tt WA 1}     & $N_{\rm H1} (\pcmsq)$ & & & $4.221 \times 10^{22}$ &
$1.995 \times 10^{22}$ & $4.172 \times 10^{22}$ \\
               & $\log \xi_1$ & & & $0.837 \pm 0.493$ & $1.762 \pm
               0.085$ & $1.663 \pm 0.033$ \\
\hline
{\tt WA 2}     & $N_{\rm H2} (\pcmsq)$ & & & & $3.489 \times 10^{23}$
& $2.187 \times 10^{23}$\\
               & $\log \xi_2$ & & & & $3.898 \pm 0.17$ & $3.454
               \pm 0.085$ \\
\hline
{\tt Fe edge}  & $\log N_{\rm Fe} (\pcmsq)$ & & & & $17.535 \pm 0.026$ &
$17.681 \pm 0.006$ \\
\hline
{\tt bbody}    & $kT$ ($\keV$) & & & & $9.359 \pm 0.281 \times
10^{-2}$ & \\
               & {\rm flux} ($\phpcmsqps$) & & & & $3.379 \pm 0.682 \times
               10^{-6}$ & \\
\hline
{\tt po}       & $\Gamma_{\rm po}$ & $1.907 \pm 0.007$ & $1.903 \pm 0.007$ & $1.966 \pm
0.014$ & $1.947 \pm 0.07$ & $2.090 \pm 0.007$ \\
               & {\rm flux} ($\phpcmsqps$) & $1.387 \pm 0.009 \times 10^{-2}$ & $1.363 \pm 0.01 \times
               10^{-2}$ & $ 1.303 \pm 0.063 \times 10^{-3}$ & $1.066
               \pm 0.027 \times 10^{-4}$ & $1.182 \pm 0.033 \times
               10^{-4}$ \\
\hline
{\tt kerrdisk}     & E ($\keV$) & & $6.576 \pm 0.037$ & $6.497
\pm 0.043$ & $6.400 \pm 0.056$ &  \\
{\tt kerrconv} & $\alpha_1$ & & $5.652 \pm 0.629$ & $5.461 \pm 1.127$
& $6.555 \pm 0.428$ & $6.057 \pm 0.255$ \\
               & $\alpha_2$ & & $2.863 \pm 0.151$ & $2.655 \pm 0.255$
               & $2.439 \pm 0.265$ & $2.776 \pm 0.183$ \\
               & $r_{\rm br}$ ($r_{\rm g}$) & & $5.747 \pm 0.816$ &
               $5.679 \pm 1.622$ & $5.139 \pm 0.867$ & $5.555 \pm 0.662$ \\
               & $a$ & & $0.970^{+0.003}_{-0.015}$ & $0.997^{+0.001}_{-0.035}$ &
               $0.997 \pm 0.001$ & $0.989^{+0.009}_{-0.002}$ \\
               & $i$ ($\degmark$) & & $20.405 \pm 2.236$ & $23.800 \pm
               1.906$ & $29.700 \pm 1.498$ & $29.699 \pm 0.709$ \\
               & $r_{\rm min}$ ($r_{\rm g}$) & & $1.738$ & $1.278$ &
               $1.278$ & $1.615 \pm 0.074$ \\
               & $r_{\rm max}$ ($r_{\rm g}$) & & $113.260 \pm 1.692$ & $102.109
               \pm 1.178$ & $134.060 \pm 0.638$ & $397.028 \pm 494.559$ \\
               & {\rm flux} ($\phpcmsqps$) & & $2.980 \pm 0.359 \times
               10^{-4}$ & $1.781 \pm 0.332 \times 10^{-5}$ & $2.677
               \pm 0.515 \times 10^{-6}$ &  \\
\hline
{\tt refl}     & {\rm Fe/solar} & & & & & $10.000 \pm 0.713$ \\
               & $\log \xi_{\rm refl}$ & & & & & $2.027 \pm 0.001$ \\
               & $\Gamma_{\rm refl}$ & & & & & $2.090 \pm 0.006$ \\
               & {\rm flux} ($\phpcmsqps$) & & & & & $1.512 \pm 0.317
               \times 10^{-7}$ \\
\hline
{\bf $\chi^2/{\rm dof}$} & & $4577/1106$ ($4.139$) & $960/1096$
($0.876$) & $934/1094$ ($0.854$) & $1742/1375$ ($1.267$) & $1793/1374$ ($1.305$) \\
\hline \hline 
\end{tabular}
\end{scriptsize}
\end{center}
\label{tab:tab1.tex}
\small{Models~$1-3$ are from $2-10 \keV$, Models~$4-5$ also include energies
from $0.6-2.0 \keV$.  Error bars quoted are all at the $1\sigma$
level except for those on the spin parameter, which are all at $90\%$
confidence.  Errors for the
column densities of the warm absorbers cannot be well constrained due
to their magnitudes.  For Model~5, the {\tt kerrconv} component
has no line energy or flux parameters, and the large error bars on
the maximum disk radius indicates that this parameter is not robustly
constrained.} 
\end{table}
}
\clearpage
\setlength{\voffset}{0mm}
\begin{figure*}[h]
\centerline{
\includegraphics[width=0.67\textwidth,angle=270]{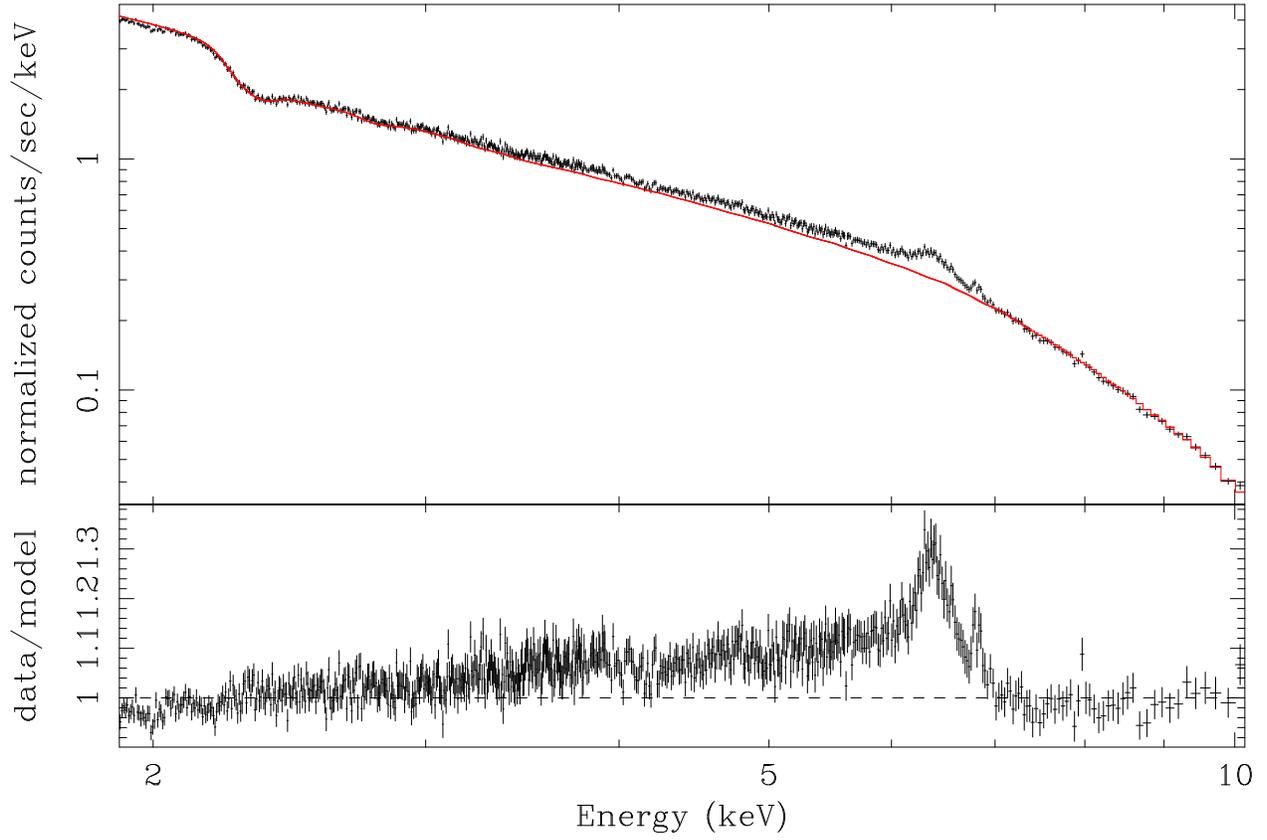}
}
\label{fig:model1}
\caption{The {\tt phabs(po)} fit to MCG--6-30-15.  Notice the
  significant deviations of the data from this model, especially
  around $6.4 \keV$.  $\chi^2/{\rm dof} = 4577/1106$ ($4.14$).}
\end{figure*}
\begin{figure*}[h]
\begin{center}
\includegraphics[width=0.36\textwidth,angle=270]{f5a.eps}
\includegraphics[width=0.39\textwidth,angle=90]{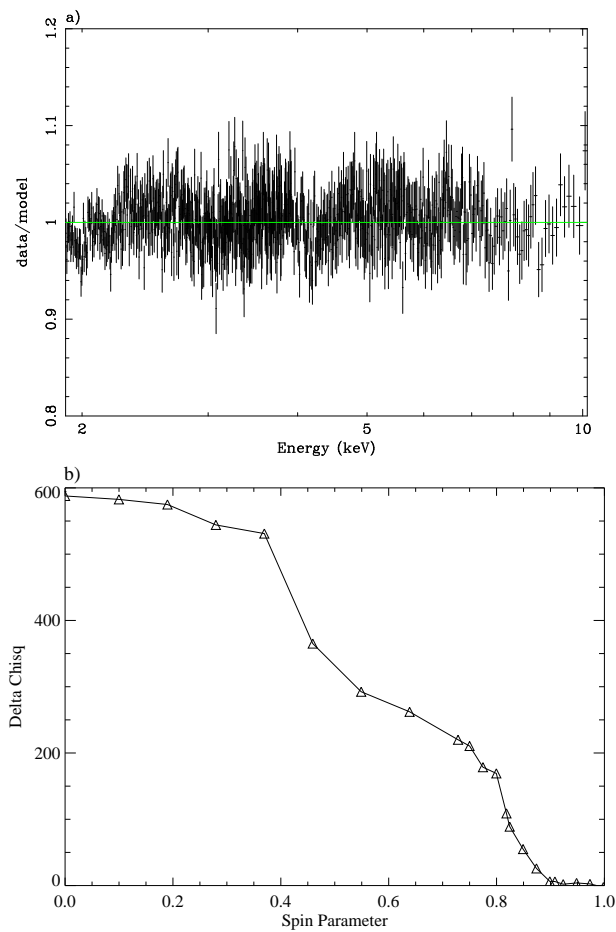}
\end{center}
\label{fig:model2}
\caption{(a) A {\tt kerrdisk} line near $6.4 \keV$ has been added to
  the {\tt phabs(po)} fit, as have two {\tt zgauss} lines modeling out
  the narrow, cold iron line at $6.4 \keV$ as well as the narrow,
ionized iron line at $6.9 \keV$.  $\chi^2/{\rm dof} = 960/1096$
($0.88$).  Note the flatness of the ratio plot shown here as compared
with that shown above for the {\tt phabs(po)} case.
(b) The corresponding plot of the change in $\chi^2$ vs. $a$
for this model.  At $90\%$ confidence, $a = 0.970^{+0.003}_{-0.015}$.}
\end{figure*}
\begin{figure*}[h]
\begin{center}
\includegraphics[width=0.36\textwidth,angle=270]{f6a.eps}
\includegraphics[width=0.39\textwidth,angle=90]{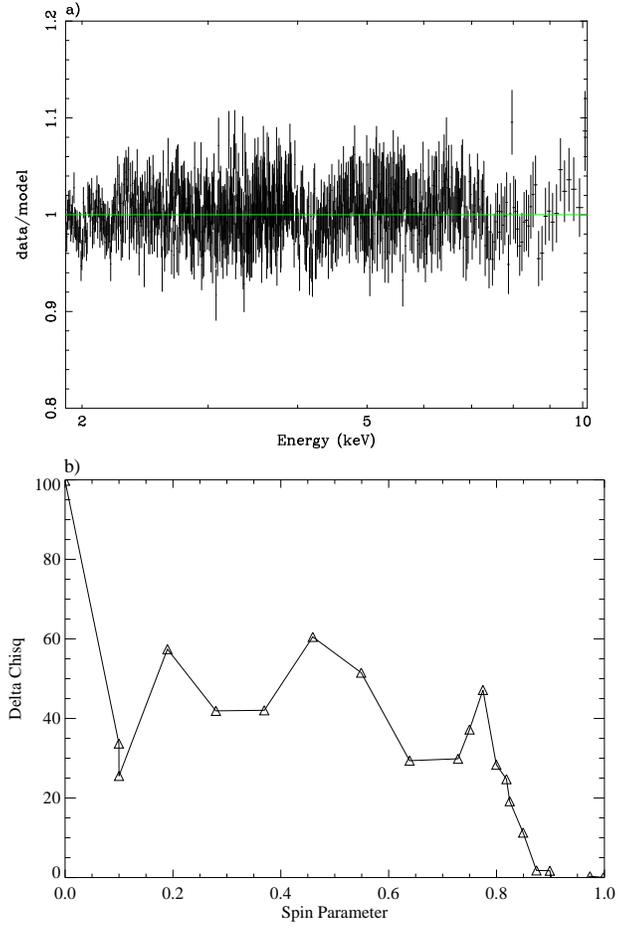}
\end{center}
\label{fig:model3}
\caption{(a) A warm absorber table has been added to the fit to try
  and model the soft end of the spectrum more accurately.
  $\chi^2/{\rm dof} = 934/1094$ ($0.85$).  (b) $\Delta\chi^2$
  vs. $a$ plot.  At $90\%$ confidence, $a = 0.997^{+0.001}_{-0.035}$.}
\end{figure*}
\begin{figure*}[h]
\begin{center}
\includegraphics[width=0.36\textwidth,angle=270]{f7a.eps}
\includegraphics[width=0.39\textwidth,angle=90]{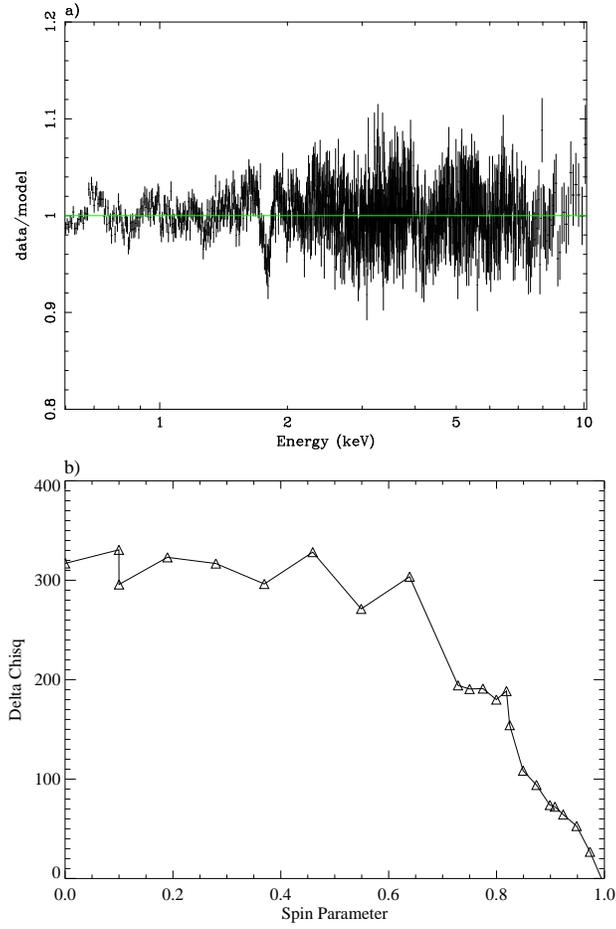}
\end{center}
\label{fig:model4}
\caption{(a) A second warm absorber has been added, as well as an iron
edge at $0.707 \keV$ and a blackbody component to model direct disk
emission.  The data now include energies from $0.6-2.0 \keV$.  The 
residual feature at $\sim 1.8 \keV$ is a calibration
  artifact (the Si K edge).  
$\chi^2/{\rm dof} = 1742/1375$ ($1.27$).  (b) $\Delta\chi^2$
vs. $a$ plot.  At $90\%$ confidence, $a = 0.997 \pm 0.001$.}
\end{figure*}

\begin{figure*}[h]
\begin{center}
\includegraphics[width=0.36\textwidth,angle=270]{f8a.eps}
\includegraphics[width=0.39\textwidth,angle=90]{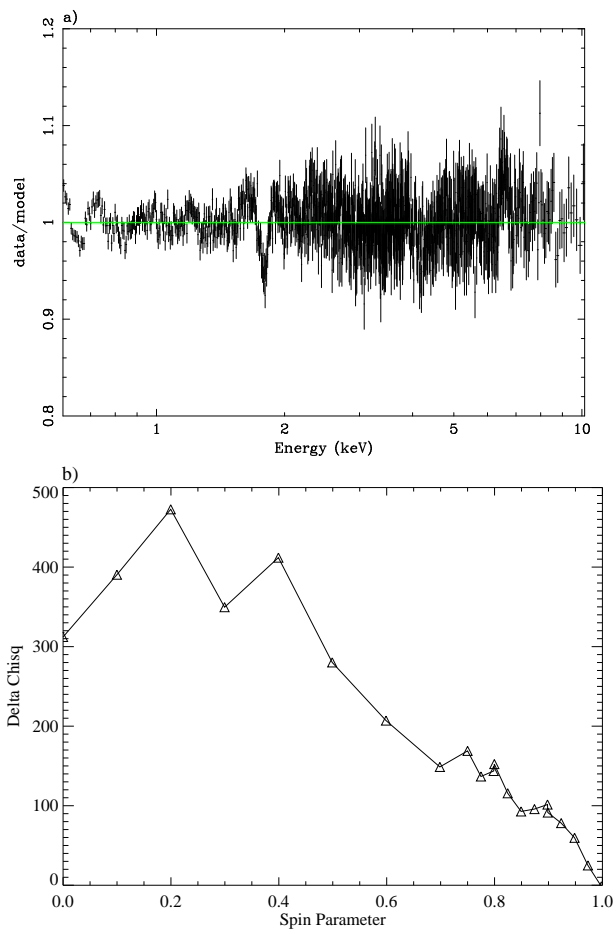}
\end{center}
\label{fig:model5}
\caption{(a) The {\tt bbody} component has been replaced by a smeared
  ionized disk reflection spectrum.  $\chi^2/{\rm dof} = 1793/1374$
  ($1.30$).  Again, the residual feature at $\sim 1.8 \keV$ is a calibration
  artifact (the Si K edge).  While this fit statistically seems less robust
  than that achieved with Model~4, the ionized reflection model is
  thought to be more physically accurate in its ability to account for
  both the soft excess in emission as well as the broad iron feature
  at $6.4 \keV$.  
(b) $\Delta\chi^2$ vs. $a$ plot.  At $90\%$ confidence, 
$a = 0.989^{+0.009}_{-0.002}$.}
\end{figure*}

\begin{figure*}[h]
\centerline{
\includegraphics[width=0.67\textwidth,angle=270]{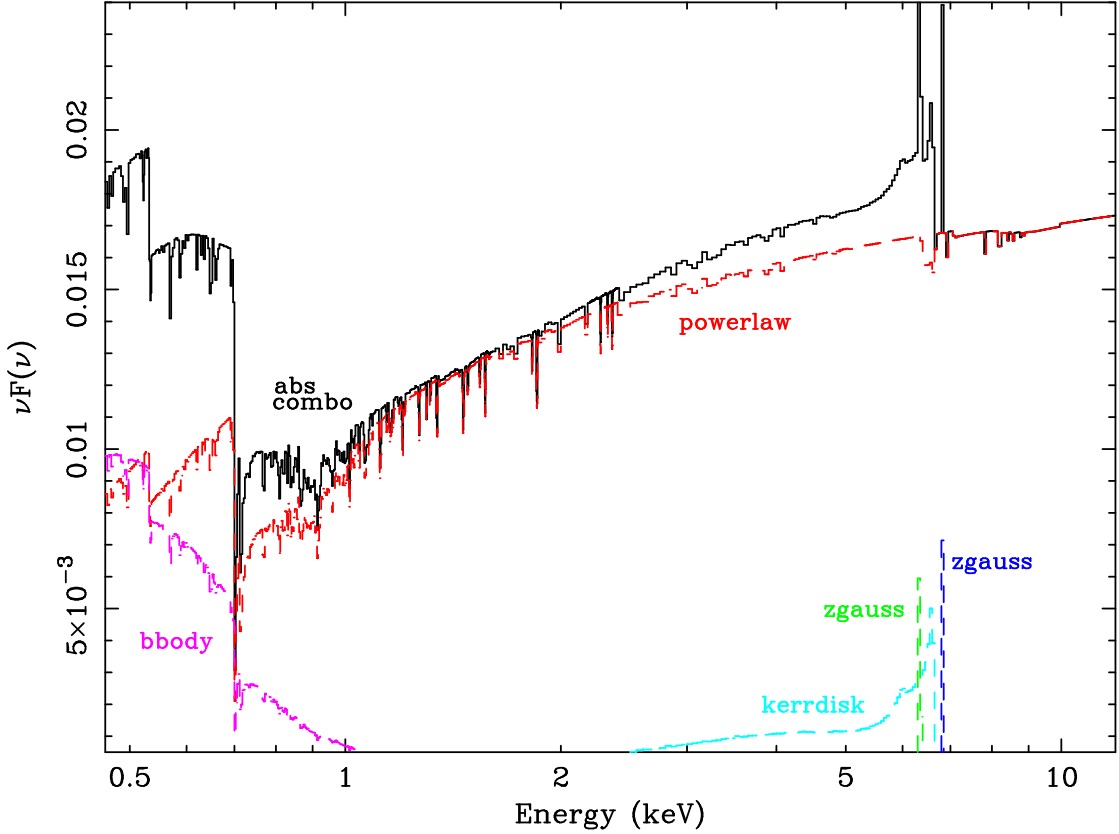}
}
\label{fig:nufnu1}
\caption{A $\nu F_{\nu}$ plot of the relative contributions of the
  model components for 
Model~4, shown
  above in Fig.~7a.  The two redshifted gaussian components ({\tt
    zgauss}) are shown
in green and dark blue, the {\tt kerrdisk} line is in light blue, the
blackbody component ({\tt bbody}) representing soft emission from the disk is in
purple, and the power-law continuum ({\tt powerlaw}) is in red.  The
soft components labeled ``abs combo'' in black represent the
combination of absorption features present in the spectrum: Galactic
photoabsorption, two warm absorber models and an iron absorption
edge.  The solid black line represents the combined model including
all the dashed components.}
\end{figure*}

\begin{figure*}[h]
\centerline{
\includegraphics[width=0.67\textwidth,angle=270]{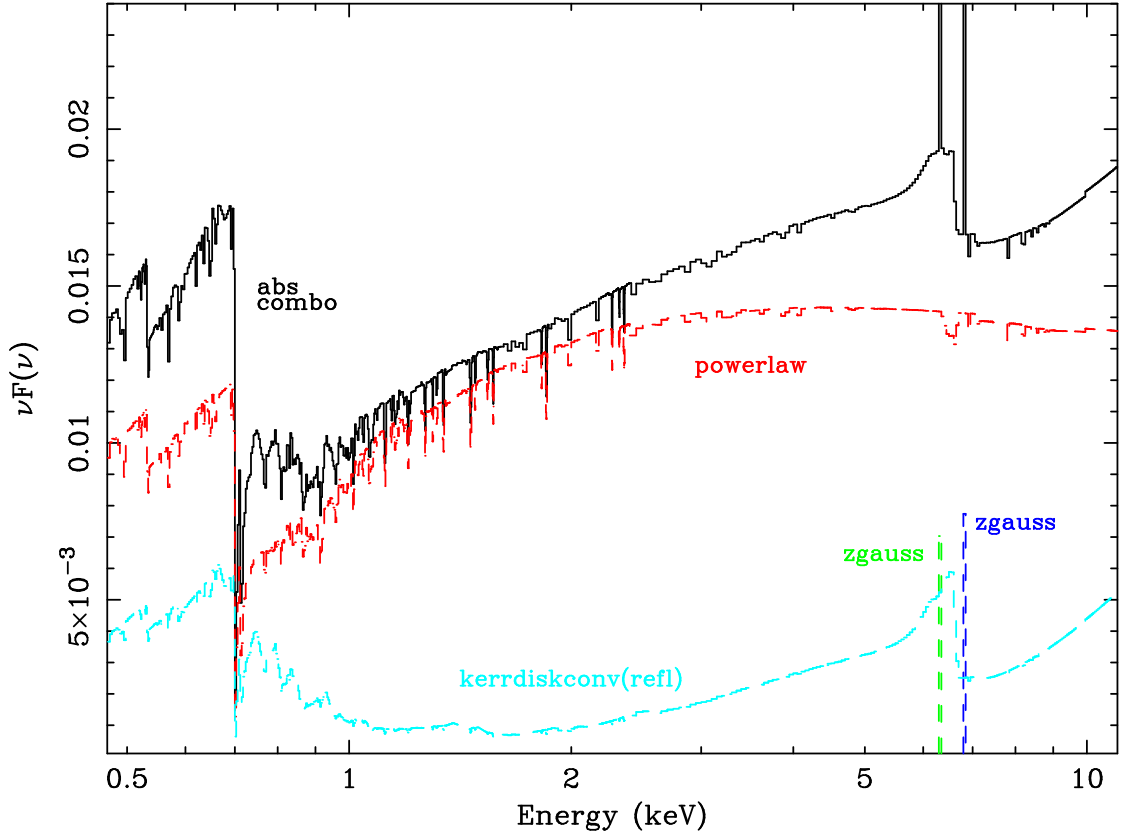}
}
\label{fig:nufnu2}
\caption{A $\nu F_{\nu}$ plot of the relative contributions of the
  model components for
  MCG--6-30-15, as above in Fig.~9.  Now the fit is
  from Model~5, as shown in Fig.~8a. The color scheme is the same as
  for Fig.~11, but in this case the blackbody component has been
  replaced by an ionized reflection spectrum to model the soft
  emission as well as the broad neutral iron line.  This component
  appears in light blue.}
\end{figure*}
\clearpage

\subsection{Ruling out a Schwarzchild Black Hole}
\label{sec:schw}

As mentioned above, Dov\v{c}iak \etal (2004) and Beckwith \& Done (2004)
argue that broad lines cannot be used as black hole spin diagnostics
due to the degeneracy that exists between the physical parameters that
go into composing the line profile.  We contend that this is not
necessarily the case; broad lines {\it can} be used to constrain black
hole spin if one takes into account the physical realism of the
best-fit parameters.  The degeneracy between parameters makes it
difficult to calculate the precise angular momentum for a given black
hole, but we can nonetheless statistically rule out certain regions of
parameter space provided that the data used has the spectral
resolution to enable accurate model fitting.

In the case of MCG--6-30-15, the Fabian \etal (2002) data set is
noteworthy for its length and unprecedented resolution of the broad
iron feature.  This makes it an ideal candidate for examining the
parameter space of the new models in question.  The width of the iron
line implies that this feature is produced in the accretion disk
immediately surrounding a rapidly spinning black hole, and indeed the
best-fit {\tt kerrdisk} parameters for the simplest model including an
iron line (Model~2, from $2-10 \keV$) suggest that $a =
0.970^{+0.003}_{-0.015}$ with $90\%$ confidence (see Fig.~5a).  The
{\tt kerrdisk} parameters are consistent with those of the {\tt
kyrline} best fit as well.  Given that the fits excluding emission
from within the radius of marginal stability imply a near-maximal spin
for the black hole, here we pose the question ``can we rule out the
non-spinning case if we relax the restriction of no emission from
within the radius of marginal stability?''.  

To answer this question we utilize the {\tt kyrline} model, since {\tt
kerrdisk} does not yet include emission from within the ISCO.  In
previous fits to MCG--6-30-15 using low-spin black hole models, it has
been found that the fit demands an inner emission radius well within
the radius of marginal stability (Dov\v{c}iak \etal 2004).
Substituting two {\tt kyrline} model components in for the {\tt
kerrdisk} component in the $2-10 \keV$ fit, we freeze $a = 0.0$ and
re-fit the data as in Model~2.  Two {\tt kyrline} components are used
because the publicly released version of {\tt kyrline} does not
support a broken power-law emissivity index for the disk (as does {\tt
kerrdisk}), so we must divide the disk up into two effective regions:
one extending outwards from $r_{\rm ms}$, and one interior to $r_{\rm
ms}$ representing the plunging region.  Because the plunging region is
physically distinct from the disk proper, any lack of continuity
between either the emissivity indices or the fluxes of the two
components is not considered problematic.

The simplest {\tt kyrline} fit for a Schwarzchild black hole is based
on Model~2.  Visually it does not differ perceptibly from the {\tt
kerrdisk} best fit for Model~2, and statistically it is a slightly
better fit: $\Delta\chi^2/\Delta{\rm dof} = -15/0$ between the two
fits.  Most notable about the fit, however, is that when we force $a =
0.0$, the fit demands an inner radius that is deep within the plunging
region ($r_{\rm min}=3.43 \pm 0.19 \,r_{\rm g}$ as compared with
$r_{\rm ms}=6 \,r_{\rm g}$ for a Schwarzchild black hole) with an
extremely high inner emissivity index $\alpha_1 = 9.08 \pm 1.36$.
Within the plunging region of a Schwarzschild black hole, we expect
the radial component of the 4-velocity to be
\begin{equation}
u^r=-c\sqrt{\frac{8}{9}-\left(1-\frac{2}{r}\right)\left(1+\frac{12}{r^2}\right)}
\end{equation}
where, here, $r$ is measured in units of $r_{\rm g}=GM/c^2$ (Reynolds \&
Begelman 1997).  For $r \approx 3.43$ we have $u^r \approx -0.22c$,
i.e., the material is already inflowing at mildly relativistic
velocities. Hence, conservation of baryon number demands that this
part of the accretion flow be extremely tenuous and, given that by
assumption it is subjected to an intense X-ray irradiation, this
material must be fully ionized (see discussion in Reynolds \& Begelman
1997 and Young, Ross \& Fabian 1998).  More quantitatively, the
analysis of Reynolds \& Begelman (1997; see Fig.~3 of this paper) show
that the ionization parameter of the accreting matter at this radius
will exceed $\log \xi=4$ for any reasonable accretion efficiency; this
is essentially fully ionized and will not imprint any obvious atomic
signatures on the backscattered spectrum.  It is therefore extremely
unlikely that a disk with such a steep emissivity profile and an inner
radius so deep within the plunging region is an accurate physical
model of the real system.

We have also performed a Schwarzchild fit to the data based on the
more complex best-fitting model from $0.6-10 \keV$ (Model~5).  Recall
that in this case we have introduced an ionized disk reflection
component to the fit, which serves the dual purpose of modeling the
soft excess and the broad Fe-K$\alpha$ line at $6.4 \keV$.  Whereas
before in Model~5 we convolved our reflection spectrum with a {\tt
kerrconv} component, here we use a {\tt kyconv} model instead
because we are demanding that $a=0.0$ in this case, and anticipate
that a large fraction of the emission will come from within the ISCO,
as was the case for our Schwarzchild fit based on Model~2 above.
Following a similar approach outlined above, we use two {\tt kerrconv}
components to allow for a broken power-law emissivity index in the
disk.  Qualitatively similar results are obtained; the inner radius of
the X-ray reflection is such a model is at $r_{\rm min}=3.02 \pm 0.31
\,r_{\rm g}$ with an extremely steep inner emissivity index
($\alpha_1=15.00 \pm 3.12$).  As before with the simpler case of a
Schwarzchild fit to Model~2, this implies that the vast majority of
the emission of this model component originates well inside the ISCO,
which is not physically realistic.  The ionization parameter for the
reflector has also risen for the Schwarzchild case from $\xi_{\rm refl}=106.36
\pm 0.94$ to $\xi_{\rm refl}=182.91 \pm 13.58$.  The iron abundance has remained
very high at ${\rm Fe/solar}=10.0 \pm 0.23$, as was the case in
Model~5.  Even with these adjustments in parameters, however, the
Schwarzchild fit visually fails to account for the entire breadth and
shape of the $6.4 \keV$ iron line.  In this case $\chi^2/{\rm
dof}=2855/1377$ ($2.08$), as compared with $1793/1374$ ($1.35$) for
the Model~5 fit where black hole spin is a free parameter.

Based on these arguments for Model~2 and Model~5 (the best-fitting
cases for $2-10 \keV$ and $0.6-10 \keV$, respectively) we can make a
strong case that a non-rotating black hole cannot viably produce the
broad iron feature in MCG--6-30-15.

\section{Discussion}
\label{sec:summary}

\subsection{Summary of Results}

We have created a new model within {\sc xspec}, called {\tt kerrdisk},
which synthesizes fluorescent emission lines produced in the accretion
disks surrounding black holes.  Our model differs from the two most
commonly used models ({\tt diskline} and {\tt laor}) in that it is
fully relativistic and allows for the spin of the black hole to be a
free parameter.  We also enable the disk emissivity index to be
modeled with greater precision as a broken power-law.  While our model
is comparable to the new models of Dov\v{c}iak \etal (2004), Beckwith
\& Done (2004) and \v{C}ade\v{z} \& Calvani (2005), we do not
pre-calculate extensive tables of the photon transfer function in the
manner of the first two sets of authors.  Rather, we pre-tabulate a
modest $20\times20$ grid of slowly varying Cunningham transfer
functions in $(a,i)$ space over a given range of radii and redshifts
within the disk, then linearly interpolate over these values to
calculate the transfer function for a given point in $(r,g^*;a,i)$
space.  The resulting line profile is indistinguishable from one
produced by performing the relativistic calculations on the fly, in
terms of accuracy, and is much faster.  Making use of linear
interpolation also negates the need for referring to gigabyte-sized
tables (e.g. Dov\v{c}iak \etal and Beckwith \& Done), making our code
more portable.

In fitting the hard spectrum of MCG--6-30-15 with the {\tt kerrdisk}
model, we have shown that the data prefer a fit with a spin parameter
that tends towards the maximum value.  A non-spinning black hole can
produce a formally adequate fit (although still statistically worse
than that achieved with a free spin parameter), but further requires a
significant fraction of the X-ray reflection to originate unphysically
deep within the plunging region.  One might argue that for
flows accreting at close to the Eddington rate, the radius of
dynamical stability might be pushed to the marginally bound orbit,
rather than the marginally stable orbit (Abramowicz \etal 1990; Chen
\etal 1997).  In principle this would mean that the optically thick
part of the flow could come substantially closer to the event horizon,
resulting in significantly more broad line contribution from this
region.  However, in the case of a Schwarzchild hole the marginally
bound orbit is still at a radius of $r_{\rm mb}=4 \,r_{\rm g}$, which
is well outside the minimum radius of emission we get for the
Schwarzchild fit based on Model~5, where $r_{\rm min}=3.02 \pm 0.31
\,r_{\rm g}$.  Also, it is unlikely that MCG--6-30-15 is accreting
close to the Eddington rate.

In fact, for the rapidly-spinning black hole preferred by our spectral
fitting, the issue of unmodeled emission from within the radius of
marginal stability diminishes in importance.  For such rapidly
spinning black holes, the geometric area of the disk within this
radius is small and the gravitational redshift of this region is
extreme.  Hence, we suspect that the inclusion of reasonable amounts
of emission from within the radius of marginal stability will have
negligible impact on our bet fitting spectral parameters, including
the spin of the black hole.

The best-fitting model for the full $0.6-10 \keV$ data set appears
to be Model~5.  The hard spectrum is best represented by a power-law
continuum and two gaussian features of iron at $6.4$ (neutral) and $6.9
\keV$ (ionized).  The soft portion of the spectrum below $\sim 2 \keV$ is well
fit by relativistically blurred emission reflected from the surface of
an ionized disk that is modified by a two-zone warm absorber, an iron
absorption edge, and Galactic photoabsorption.  This ionized emission
also contributes to the breadth and morphology of the observed iron
feature near $6.4 \keV$.  As described above in \S\ref{sec:ion_disk},
this model yields a best-fit spin parameter for the black hole in
MCG--6-30-15 of $a = 0.989^{+0.009}_{-0.002}$.  Due to the aforementioned
degeneracies in the broad iron line model parameters
(\S\ref{sec:intro}) this value should not be interpreted as exact, but
the fact that near maximal spin is approached in each of the
Models~2-5 strongly indicates that the data prefer a rapidly spinning
black hole.

\section{Conclusions and Future Work}
\label{sec:conclusions}

Both Dov\v{c}iak \etal (2004) and Beckwith \& Done (2004) make valid
points about the difficulty of accurately calculating the spin
parameter of a black hole based on broad line profiles from the disk
alone.  Given the number of parameters governing the {\tt kerrdisk}
line profile (nine, to be exact), some degeneracy between them is
certainly to be expected, and it is possible to generate statistically
indistinguishable line profiles using different combinations of
parameters.  This does not render the use of broad lines as spin
diagnostics obsolete, however: provided that one uses care in
determining whether parameter values are physically reasonable, it is
possible to at least constrain the spin parameter to be within a
certain range of values.  In the example of MCG--6-30-15 above, we
have clearly shown that the data rule out a non-spinning black
hole if we also demand that the values for the disk emissivity indices
are physically realistic.  The plots of $\Delta \chi^2$ vs. $a$
clearly show the improvement in fit achieved when one frees the spin
parameter.  In this case the fit strongly tends toward maximal spin.  

The combination of such a high spin and such a high iron abundance in
the disk may understandably give one pause when considering the
precision of our best-fitting Model~5.  Both $a$ and $Z_{\rm Fe}$ push
the upper limits of their respective parameter spaces in order to try
to account for the extreme breadth and strength of the Fe-K$\alpha$
feature in the spectrum.  As we have mentioned, however, in this work
we are seeking to establish a robust method for constraining black
hole spin via X-ray spectroscopy and are not claiming that our
best-fit parameter values should be interpreted as exact.  The spin of
the black hole may indeed reach $a=0.989$, but there could also be
some contribution to the line from highly redshifted radiation
originating from within the ISCO that we have not yet accounted for,
which could reduce the need for such a high spin value.  What we can
say with confidence is that fits to the data with a Schwarzchild black
hole are not physically sound.  Likewise, the disk may indeed have
nearly ten times the solar value of iron, but underabundances in some
of the lighter elements may also simulate an effectively iron-rich
environment which could contribute to the observed width of the
Fe-K$\alpha$ line (Reynolds, Fabian \& Inoue 1995).  {\it Suzaku}
observations of this source will be invaluable for untangling the
parameters of the ionized reflection model.  Much work remains to be
done on the subjects of modeling accretion disks and isolating black
hole spins, and improvements in the former will undoubtedly help us
place more accurate constraints on the latter.

If MCG--6-30-15 is indeed a rapidly spinning black hole, as seems
likely given our spectral modeling, it is astrophysically interesting
for several reasons.  As mentioned in \S\ref{sec:intro}, rapidly
spinning holes can, in principle, experience a magnetic torque by the
fields threading the accretion disk at the radius of marginal
stability.  This torque can theoretically extract rotational energy
from the hole itself, significantly enhancing the amount of
dissipation in the inner accretion disk.  The steepest dissipation
profiles would be obtained if the magnetic torque is applied
completely at the radius of marginal stability (Agol \& Krolik 2000).
Therefore, only for rapidly spinning black holes would one expect to
observe such a steep dissipation profile: in this scenario $r_{\rm
ms}$ is dragged inward very close to the event horizon, so the torque
is strongest here.  Such an effect would manifest itself via strongly
redshifted reflection features in the spectrum, since the strong
dissipation very close to the event horizon would mean that most of
the emission would originate from this region.  Based on its own
best-fit emission profile, MCG--6-30-15 may in fact be giving us a
glimpse of this phenomenon at work (Wilms \etal 2001).  It should be
noted, however, that interpreting the detected reflection features in
this way demands that little of the observed emission originate from
the plunging region within $r_{\rm ms}$.  Taken in this context, the
relatively steep emissivity index we found for the ionized disk in our
best-fitting Model~5 ($a=0.989$, $\alpha_1=6.06$) is not unexpected,
and may be indicative of this type of magnetic torquing.

\section*{Acknowledgments}

We gratefully acknowledge support from NSF grant AST0205990.  We also
thank Andy Fabian for his helpful comments on this work.  Michael
Nowak, Sera Markoff and Andy Young provided insightful discussions
about {\sc xspec} and {\sc isis}, as well as previous X-ray fits to
MCG--6-30-15.

\end{document}